\documentclass[9.5pt,journal,compsoc]{IEEEtran}
\ifCLASSOPTIONcompsoc
  \usepackage[nocompress]{cite}
\else
  \usepackage{cite}
\fi
\usepackage{cite}
\usepackage{amsmath,amssymb,amsfonts}
\allowdisplaybreaks
\usepackage{algorithm, algorithmic}
\usepackage{graphicx}
\usepackage{textcomp}
\usepackage{xcolor}
\usepackage{optidef}
\usepackage{subcaption}
\usepackage{lipsum,graphicx}
\usepackage[font={small}]{caption}
\usepackage[english]{babel}
\usepackage{environ}         
\usepackage{etoolbox}        
\usepackage{xparse}
\usepackage{etoolbox}
\usepackage{xcolor}
\usepackage{threeparttable}  
\usepackage{footnote}  
\makesavenoteenv{table} 

\ifCLASSINFOpdf
\else
\fi

\begin{document}
\title{\Large Enhancing Spectrum Efficiency in 6G Satellite Networks: A GAIL-Powered Policy Learning via Asynchronous Federated Inverse Reinforcement Learning}

\author{Sheikh Salman Hassan,~\IEEEmembership{Member,~IEEE,} Yu Min Park,~\IEEEmembership{Student Member,~IEEE},\\ Yan Kyaw Tun~\IEEEmembership{Member,~IEEE,} Walid Saad,~\IEEEmembership{Fellow,~IEEE,} Zhu Han,~\IEEEmembership{Fellow,~IEEE,}\\
and~Choong Seon Hong,~\IEEEmembership{Fellow,~IEEE.}
\IEEEcompsocitemizethanks{\IEEEcompsocthanksitem Sheikh Salman Hassan, Yu Min Park, and Choong Seon Hong are with the Department of Computer Science and Engineering, Kyung Hee University, Yongin, 446-701, Republic of Korea.\protect\\
E-mail: \{salman0335, yumin0906, cshong\}@khu.ac.kr
\IEEEcompsocthanksitem Yan Kyaw Tun is with the Department of Electronic Systems, Aalborg University, 2450 København SV, Denmark. \protect E-mail:  ykt@es.aau.dk
\IEEEcompsocthanksitem Walid Saad is with the Department of Electrical and Computer Engineering, Virginia Tech, VA, 24061, USA. \protect
E-mail:  walids@vt.edu
\IEEEcompsocthanksitem Zhu Han is with the Department of Electrical and Computer Engineering at the University of Houston, Houston, TX 77004 USA, and also with the Department of Computer Science and Engineering, Kyung Hee University, Seoul, South Korea, 446-701. \protect\\
E-mail:  hanzhu22@gmail.com.

}
}

\markboth{Journal of \LaTeX\ Class Files,~Vol.~14, No.~8, August~2015}%
{Shell \MakeLowercase{\textit{et al.}}: Bare Advanced Demo of IEEEtran.cls for IEEE Computer Society Journals}

\IEEEtitleabstractindextext{%
\begin{abstract}
In this paper, a novel generative adversarial imitation learning (GAIL)-powered policy learning approach is proposed for optimizing beamforming, spectrum allocation, and remote user equipment (RUE) association in NTNs. Traditional reinforcement learning (RL) methods for wireless network optimization often rely on manually designed reward functions, which can require extensive parameter tuning. To overcome these limitations, we employ inverse RL (IRL), specifically leveraging the GAIL framework, to automatically learn reward functions without manual design. We augment this framework with an asynchronous federated learning approach, enabling decentralized multi-satellite systems to collaboratively derive optimal policies. The proposed method aims to maximize spectrum efficiency (SE) while meeting minimum information rate requirements for RUEs. To address the non-convex, NP-hard nature of this problem, we combine the many-to-one matching theory with a multi-agent asynchronous federated IRL (MA-AFIRL) framework. This allows agents to learn through asynchronous environmental interactions, improving training efficiency and scalability. The expert policy is generated using the Whale optimization algorithm (WOA), providing data to train the automatic reward function within GAIL. Simulation results show that the proposed MA-AFIRL method outperforms traditional RL approaches, achieving a $14.6\%$ improvement in convergence and reward value. The novel GAIL-driven policy learning establishes a novel benchmark for 6G NTN optimization.
\end{abstract}

}

\maketitle
\IEEEdisplaynontitleabstractindextext
\IEEEpeerreviewmaketitle
\ifCLASSOPTIONcompsoc
\IEEEraisesectionheading{\section{Introduction}\label{sec:introduction}}
\else
\section{Introduction}
\label{sec:introduction}
\fi
\IEEEPARstart{T}{he} vast expanses of the Earth's surface, predominantly comprising oceans and deserts, present significant challenges in establishing large-scale base stations (BSs) to address the escalating wireless communication demands and the proliferation of Internet-of-Things (IoT) devices \cite{NTN-need, Own_iot}. Non-terrestrial networks (NTNs), with a particular emphasis on satellite-based systems, are rapidly evolving as an extension of terrestrial infrastructure, leading to a substantial increase in remote user equipment (RUE) connectivity and data transmission volumes \cite{NTN+AI-motivation}. Among these, low Earth orbit (LEO) satellite communication networks have attracted considerable attention due to their extensive global coverage capabilities and potential for reliable communication services. Several satellite operators, including Starlink \cite{fomon2022starlink}, OneWeb \cite{oneweb2024solutions}, Amazon \cite{fcc2020kuiper}, and Boeing \cite{fcc2021boeing}, have launched or are planning to launch LEO satellite networks to serve millions of potential terrestrial terminals. The lower orbital altitude of LEO satellites\footnote{Hereinafter, the term \textit{LEO satellites} will be used interchangeably with \textit{satellites} unless explicitly stated otherwise.} results in shorter transmission delays and reduced path loss compared to geostationary Earth orbit (GEO) satellites. Additionally, a constellation of multiple satellites can achieve global coverage. Moreover, technologies such as cellular communication, multiple access, point beam, and frequency multiplexing provide robust technical support for LEO satellite communications. 

The growing demand for dynamic spectrum resources and the surge in remote user equipments (RUEs) access pose significant challenges to satellite communication networks. To address these challenges, several advanced access schemes have been proposed, including dynamic spectrum access (DSA), non-orthogonal multiple access (NOMA), cognitive radio (CR), and the use of multiple spot beams, where DSA leverages deep learning for resource allocation, which requires considerable computational power \cite{DSA}. Similarly, NOMA, when integrated with orthogonal frequency division multiplexing (OFDM), enhances spectrum efficiency \cite{NOMA}. Additionally, CR systems improve spectrum utilization by adjusting transmission parameters based on real-time environmental sensing \cite{CR}. Finally, the multiple spot beam technique expands satellite antenna coverage, although it may introduce beam interference issues \cite{multibeam}. However, to ensure quality of service (QoS) for RUEs while optimizing spectrum efficiency (SE), new network management strategies are imperative. Traditionally, model-driven approaches have been employed to optimize parameters such as beamforming, user association, and spectrum allocation. However, these approaches are often inadequate in dynamic environments characterized by rapid channel fading. Their reliance on repeated iterations and precise channel models renders them computationally intensive and limits their adaptability in real-world applications. Therefore, more efficient, adaptive techniques are required to handle the complexities of modern satellite communication networks.

Recent advancements have seen a rise in prominence for self-learning techniques \cite{saad2024artificialgeneralintelligenceaginative}, particularly reinforcement learning (RL), in the domains of DSA and spectrum sensing. RL enables agents to learn optimal strategies through interactions with their environment (e.g., see \cite{DRL-NTN-1, DRL-NTN-2} and \cite{DRL-NTN-3}). However, a key challenge in RL lies in designing an effective reward function for the Markov decision process (MDP) that captures desired objectives while considering the problem constraints. The manual design of the reward function which is composed of objective functions and requires constraints is a complex process and it needs a trial-and-error and time-consuming parameter-tuning process, with limited generalizability to diverse network scenarios. Moreover, optimizing resource allocation in NTNs presents significant challenges. The traditional RL techniques depend on manually designed reward functions which can be susceptible to bias and require extensive simulations to learn. This limits their adaptability and performance in dynamic and heterogeneous network environments. Therefore inverse RL (IRL) offers a solution to this challenge by leveraging an artificial neural network (ANN) as the reward function \cite{NEURIPS2023_da409884}. The ANN is trained on data obtained from an expert policy, allowing it to learn and refine the reward function through interaction with the environment. Therefore IRL demonstrates effectiveness in handling cellular network management, i.e., power allocation problems as given in \cite{IRL}. Moreover, to promote cooperation among satellite nodes and enhance training efficiency while preserving privacy, federated learning (FL) is employed. During FL training, nodes upload local model parameters to a central network, which then performs global model aggregation. This fosters both training efficiency and decentralization within the distributed system.

Motivated by the above challenges of optimizing beamforming and resource allocation in decentralized satellite networks, the main contribution of this work is to develop a novel framework based on a generative adversarial imitation learning (GAIL)-powered IRL with federated learning (FL) tailored for NTNs \cite{irl-2, why-IRL}. Our solution focuses on optimizing satellite-to-RUE links, i.e., beamforming, spectrum allocation, and RUE association while maintaining centralized control over satellite-to-infrastructure (S2I) links through terrestrial gateway station (TGS) where FL enables decentralized collaboration across satellites. First, we introduce an efficient IRL-based approach to maximize spectrum efficiency (SE) under QoS constraints in a multi-user, multi-satellite environment, leveraging a time-varying channel model based on free-space path losses. To the best of our knowledge, this is the first application of IRL to address constrained optimization in NTNs. Second, we implement the GAIL framework within our IRL model to directly derive policies from expert demonstrations, where GAIL-powered policy learning enables us to overcome the limitations of manually designed reward functions, making the optimization process more autonomous and adaptable to dynamic network conditions. Third, we utilize the Whale optimization algorithm (WOA) as an expert policy, generating data to train the automatic reward function. This allows the IRL model to learn from high-quality expert data, enhancing the robustness and effectiveness of the derived policies. By incorporating asynchronous federated learning, our approach facilitates decentralized training across satellite agents, ensuring scalability and adaptability in dynamic NTN environments. This novel IRL-based method sets a new benchmark for NTN optimization, advancing the field towards autonomous and intelligent network management solutions. Our key contributions are summarized below:
\begin{itemize}
 \item We formulate an optimization problem focused on maximizing the SE of satellite networks, involving multiple RUEs and satellites. To solve this, we introduce the multi-agent asynchronous federated IRL (MA-AFIRL) algorithm, demonstrating its effectiveness in distributed problem-solving scenarios within dynamic NTNs.
\item We develop a novel IRL-based approach for resource allocation in NTNs, considering QoS constraints and leveraging a general time-varying channel model. This IRL method eliminates the need for manually designed reward functions, enabling more flexible and robust optimization.
\item Within our IRL framework, we implement the GAIL model to directly learn policies from an expert. The WOA serves as the expert policy, generating data to train the automatic reward function, which is central to our GAIL-powered learning approach.
\item We introduce an asynchronous federated learning approach that aggregates model updates to produce a global model with strong generalization capabilities. This decentralized learning framework facilitates communication and collaboration across multiple agents while transmitting and aggregating trained model parameters over the ground network. Our solution is particularly well-suited to dynamic satellite network environments, where synchronous updates are challenging, ensuring robust and adaptable performance.
\item Simulation results show that the proposed approach significantly improves the SE  significant improvements compared to centralized and traditional RL-based schemes, e.g., PPO, achieving a $14.6\%$ improvement in convergence and reward metrics across various satellite constellation configurations. Our MA-AFIRL-based approach outperforms established baselines, achieving SE with QoS guarantees in dynamic NTN environments, thereby validating the efficacy of the proposed method.
\end{itemize}

The rest of this paper is organized as follows. Section \ref{rel_work} summarized the related literature. Section \ref{sys_model} presents a comprehensive analysis of the system model, followed by a formulation of the problem in Section \ref{prob_form}. The proposed algorithm is described in Section \ref{sol_app}. Section \ref{simul} presents the simulation results, where the results are consolidated and examined. Finally, the conclusions derived from our work are summarized in Section \ref{conc}.

\section{Related Works}
\label{rel_work}
In this section, we provide an overview of the related works in NTNs' networking and communication management and RL-empowered NTN management as well as the limitations of the prior art.
\subsection{NTNs' Networks and Communication Management}
The terrestrial communication systems, though advanced, are costly and mostly serve urban areas \cite{NTN-1}. In contrast, NTNs using satellites and UAVs base stations, offer flexible, cost-effective connectivity, especially for remote regions \cite{NTN-2, NTN-3}. However, the increasing demand for applications in intelligent transportation systems (ITS) has highlighted the limitations of conventional terrestrial networks, particularly in connecting remote nodes such as airplanes and ships. For instance, the work in \cite{NTN-4} develops a new approach that leverages satellite access networks (SANs) to enable multi-access edge computing (MEC) servers to offer data offloading and computation services. Meanwhile, the authors in \cite{NTN-net-comm-1} focus on optimizing a cognitive satellite-terrestrial covert communication network by maximizing covert communication rates. This is achieved through the optimization of satellite transmit power and base station beamforming under imperfect channel state information (CSI) while ensuring achievable rate and power constraints.

Additionally, the work in \cite{NTN-net-comm-2} proposes a distributed, lightweight, and stateless core network architecture for integrating terrestrial cellular networks with LEO satellites in 6G networks. This architecture addresses challenges such as signaling storms and service disruptions by decoupling and centrally managing network function (NF) contexts, thereby ensuring seamless cooperation with terrestrial core networks. Moreover, \cite{NTN-net-comm-3} introduces an energy-efficient multi-connectivity (MC) framework for uplink communications in multi-orbit NTNs. This framework allows user terminals to connect simultaneously to multiple satellites, enhancing throughput while minimizing energy consumption. Finally, the study in \cite{NTN-net-comm-4} explores the performance of two-way satellite-high altitude platform (HAP)-terrestrial networks using NOMA to improve spectrum efficiency and connectivity. The research specifically examines the impact of CSI information and successive interference cancellation on network performance. 

\subsection{RL-Empowered NTNs Management}
There have been a number of recent works that leverage RL to address various NTN problems.  For example, the work in \cite{RL+NTN-NEW-1} introduces a UE-driven deep RL (DRL) approach, where a centralized agent located at the backhaul of NTN base stations (NTN-BSs) trains a deep Q-network (DQN). This model enables UEs to independently make access decisions based on the learned parameters from the DQN, enhancing network efficiency. The authors in \cite{RL+NTN-NEW-2} propose a resource allocation framework for terrestrial-satellite networks based on NOMA. This framework integrates a local cache pool to reduce latency and improve energy efficiency, employing a multi-agent deep deterministic policy gradient (MA-DDPG) method to optimize user association, power control, and cache design. The study in \cite{metaRL} proposes a distributed distribution-robust meta RL (D$^2$-RMRL) algorithm to optimize data pre-storage and routing in dynamic, resource-constrained cube satellite networks, improving pre-store hit rates and convergence speed compared to baseline methods.

Meanwhile, the work in \cite{RL+NTN-NEW-3} develops a multi-agent DRL (MA-DRL) framework for uplink channel allocation in multi-beam satellite systems. The framework is designed to minimize interference with terrestrial stations while meeting QoS requirements. To address the non-stationarity inherent in MA-DRL environments, agents are trained sequentially, improving the system's adaptability. In \cite{RL+NTN-NEW-4}, a multi-time-scale DRL scheme is proposed for radio resource optimization in NTNs, where LEO satellites and UEs collaborate through independent decision-making processes across different control cycles, optimizing network performance. The study in \cite{gan-rl} proposes an experienced DRL framework, pre-trained with generative adversarial networks (GANs), for model-free resource allocation in URLLC-6G downlink wireless networks, achieving near-optimal performance in reliability, latency, and rate while adapting quickly to extreme conditions.

Moreover, the work \cite{RL+NTN-NEW-5} presents an innovative approach for an integrated satellite-aerial-terrestrial relay network (ISATRN) using HAPs. This system employs mixed free-space optical (FSO)/radio frequency (RF) transmissions and UAVs equipped with reconfigurable intelligent surfaces (RISs). The study optimizes the system's ergodic rate through joint optimization of UAV trajectory, RIS phase shifts, and active transmit beamforming using an energy-efficient DRL approach with an enhanced long short-term memory (LSTM) double DQN (DDQN) framework. Finally, \cite{RL-NTN-net-comm-2} addresses interference management in NTNs via LEO satellites, developing an MA-RL framework for multi-beam uplink channel allocation. This framework prioritizes actions based on interference levels, improving learning efficiency and minimizing interference with terrestrial stations under QoS constraints. Additionally, \cite{NTN+RL+JSAC} tackles challenges such as path loss and data routing between satellites and remote UEs using an MA-RL approach based on proximal policy optimization (PPO). 

\subsection{Limitations of the Prior Art}
Despite the NTN's potential in 6G networks, it still faces key challenges such as long propagation delays, complex resource allocation, and dynamic handover management. These issues are exacerbated by the difficulty of obtaining complete environmental information, making efficient decision-making critical for optimal network performance. While the mentioned works \cite{NTN-1, NTN-2, NTN-3, NTN-4, NTN-net-comm-1, NTN-net-comm-2, NTN-net-comm-3, NTN-net-comm-4} present effective solutions for NTNs, they mainly overlook the dynamic and complex nature of spectrum allocation, beamforming, and user association in NTN environments which limits their effectiveness in optimizing network resource allocation.  Moreover, RL-based approaches \cite{RL+NTN-NEW-1, RL+NTN-NEW-2, RL+NTN-NEW-3, RL+NTN-NEW-4, RL+NTN-NEW-5, RL-NTN-net-comm-2, NTN+RL+JSAC} have shown promise in optimizing NTNs, however, they struggle with scalability, non-stationarity, and the high-dimensional state-action space in dynamic NTN environments. Therefore, To address NTN challenges, we propose a novel MA-AFIRL framework that combines IRL with FL for optimizing beamforming, spectrum allocation, and RUE association, where IRL enhances decision-making by learning reward functions from expert behavior, while FL enables decentralized, privacy-preserving collaboration across satellites. Additionally, an asynchronous update system allows satellites to refine strategies in real time, improving resource allocation and network optimization in dynamic NTN environments.

\section{System Model and Problem Formulation}
\label{sys_model}
\subsection{Network Model}
\begin{figure}[t]
        \centering
        \captionsetup{justification=centering,singlelinecheck=false}
        \includegraphics[width=0.9\columnwidth]{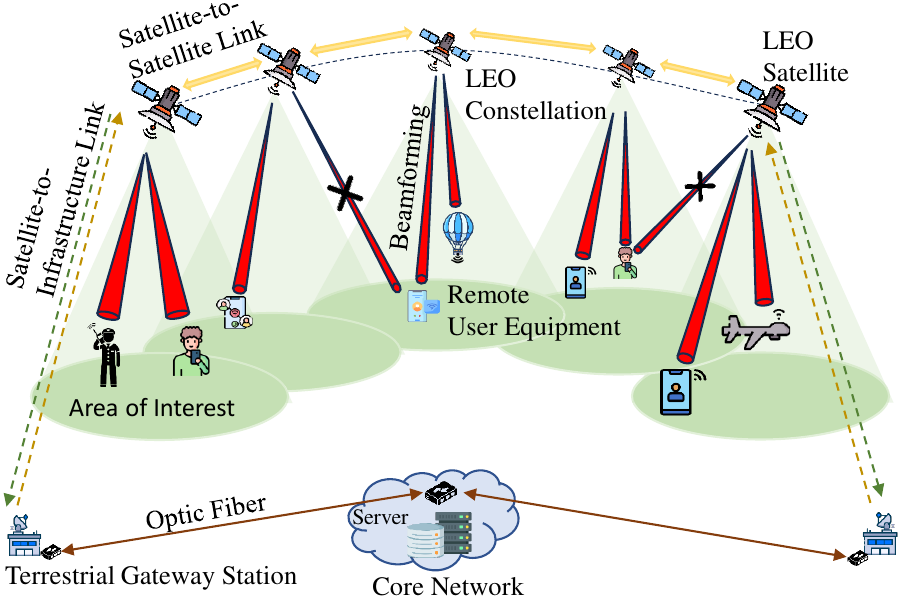}
        \caption{Multi-satellites architecture for joint beamforming, spectrum allocation, and RUE scheduling management.}
        \label{system_model}
    \end{figure}

As shown in Fig. \ref{system_model}, we consider an NTN\footnote{In this work, NTNs are considered satellite-based networks.} system composed of three components: a set $\mathcal{S}$ of $S$ LEO satellites, a set $\mathcal{P}$ of $P$ terrestrial gateway stations (TGSs), and a set $\mathcal{U}$ of $U$ RUEs. We consider downlink transmissions with the satellites collaboratively serving multiple RUEs. The NTN's components are interconnected with the core network, where communication exchange between satellites and TGSs uses satellite-to-infrastructure (S2I) links, necessitating high data rates for efficient data routing. Additionally, satellites communicate directly with each other through optical satellite-to-satellite (S2S) links, also called inter-satellite links (ISLs), when proximal TGSs are unavailable and to cooperatively serve in the constellation. The S2S links support functions such as handover, relaying, and traffic load balancing. 

All satellites are assumed to possess regenerative payloads and operate within a LEO constellation, and they use the Ka-band with full frequency reuse (FFR). Optical S2S links enable satellite data exchange, with onboard distributed computing capabilities facilitating computation load sharing. However, the TGSs are positioned statically within the satellite coverage area and periodically transmit constellation topological relationships to satellites via line-of-sight (LoS). In contrast, satellites exchange topological information through S2S links. Moreover, the satellites maintain connectivity with at least one TGS or neighboring satellites during each time interval to receive network management updates despite their periodic positions within LEO constellations. Given the emphasis on dynamic beamforming and RUEs link resource management, previous research has already studied constellation design optimization to mitigate redundancy \cite{LEO_constellation}. The initial satellite positioning is configured over areas of interest with the highest demand for SANs.

\subsection{Communication Channel Model}
We adopt a heterogeneous antenna configuration, where GUs are equipped with single-antenna very small aperture terminals (VSATs), while the satellites use a uniform planar array (UPA) directed towards the Earth. Each satellite is equipped with a large-scale UPA consisting of $N_s = N^x_s \times N^y_s$ antenna elements, where $N^x_s$ represents the number of elements on the $x$-axis and $N^y_s$ is the number of elements on the $y$-axis. The phase of each antenna element can be digitally controlled. In this scenario, the x-axis represents the direction of the satellite's movement, while the y-axis is the orthogonal direction of the movement. In this system, multiple satellites cooperatively provide communication services for their RUEs utilizing FFR. A single satellite can support up to $N^{\textrm{beam}}_s$ concurrent RUEs by generating $N^{\textrm{beam}}_s$ independent spot beams, each serving a unique RUE at any given time. Consequently, each RUE experiences co-channel interference from all visible satellites (i.e., when regions overlap). Therefore when signals are received, the RUE antenna focuses (i.e., directs its main lobe) toward the designated serving satellite which results in the intended signal and interference from this satellite experiencing the maximum gain of the RUE antenna.

Conversely, the strength of the interfering and intended signals originating from other visible satellites is attenuated due to the off-boresight angles of these satellite-RUE links and the inherent narrow-beam characteristics of the considered VSAT. It is important to note that while the set of visible satellites for each RUE is known a priori, the specific active links between satellites and RUEs remain unknown and require determination. We introduce a discrete binary variable denoted by $\chi_{su}$ to address this. This variable indicates the link association state between satellite s and RUE u, where $\chi_{su}$ = $1$ signifies that satellite s is actively serving RUE u, and $\chi_{su}$ = 0 indicates the opposite.

Our study leverages the technical reports of the 3rd Generation Partnership Project (3GPP) and the International Telecommunication Union Radio-communication Sector (ITU-R) to model the propagation channel \cite{3GPPNRtoSupportNonTerrestrialNetworks, ITUAttenuationAtmosphericGases, LEO_channel_modeling_LOS, ITUPropDataEarthSpaceLandMobile, LEO_channel_modeling, 3GPPUplinkChannelModel }. We focus on a scenario with no rain and cloud attenuation, where all RUEs are distributed in a suburban area. Under these assumptions, the multiple-input single-output (MISO) channel response between satellite $s$ and RUE $u$ at time $t$ and frequency $f$ can be modeled as:
\begin{equation}
    h_{s,u} [t,f] = g_{s,u} \cdot \exp\{\Bar{j}2\pi [t\nu_{s,u} - f\tau_{s,u}] \},  \label{channel_model}
\end{equation}
where $g_{s,u}$ is the channel gain coefficient, $\nu_{s,u}$ Doppler shift in both links due to the satellite movement, and $\tau_{s,u}$ are the propagation delay. The channel gain coefficient, $g_{s,u}$ incorporates large-scale path loss (PL) and antenna gains. It can be mathematically expressed as:
\begin{equation}
    g_{s,u} = \sqrt{G_s G_u} \cdot 10^{-\frac{1}{10} \textrm{PL[dB]} },
\end{equation}
where $G_s$ represents the gain of the satellite UPA and $G_u$ represents the gain of the antenna at RUE $u$ for the satellite $s$ link. The large-scale path loss (PL) experienced between a satellite and a RUE is calculated as the sum of three terms as follows \cite{3GPPNRtoSupportNonTerrestrialNetworks}:
\begin{equation}
    \textrm{PL}[\textrm{dB}] = \textrm{PL}_b[\textrm{dB}] + \textrm{PL}_g[\textrm{dB}] + \textrm{PL}_s[\textrm{dB}],
\end{equation}
where $\textrm{PL}_b$ represents the large-scale path loss due to free-space propagation, $\textrm{PL}_g$ captures the attenuation caused by atmospheric gasses, and $\textrm{PL}_s$ accounts for the scintillation effects arising from either the ionosphere or troposphere. Each of these terms is defined as follows:
\begin{enumerate}
    \item \textbf{Basic path loss ($\textrm{PL}_b$)}: This represents the attenuation due to free space propagation and is expressed as:
    \begin{equation}
        \textrm{PL}_b[\textrm{dB}] = \textrm{FSPL}(d_0, f_c)[\textrm{dB}] + \textrm{SF}[\textrm{dB}] + \textrm{CL}[\textrm{dB}], 
    \end{equation}
    where $\textrm{FSPL}(d_0, fc)$ denotes the free-space path loss dependent on the transmission distance $d_0$ and carrier frequency $f_c$. SF accounts for shadow fading modeled by a log-normal distribution. $\textrm{CL}$ represents clutter loss, which is typically negligible due to the assumed Line-of-Sight (LoS) condition and is set to $0$ dB.
    \item \textbf{Atmospheric gas attenuation ($\textrm{PL}_g$)}: This attenuation depends primarily on frequency, elevation angle, altitude above sea level, and water vapor density. The specific model is adopted from \cite{ITUAttenuationAtmosphericGases}.
    \item \textbf{Scintillation attenuation ($\textrm{PL}_s$)}: In the Ka-band, ionospheric scintillation is negligible. Therefore, $\textrm{PL}_s$ only accounts for tropospheric scintillation effects.
\end{enumerate}
\subsection{Communication Link Analysis}
The satellite downlink transmission to RUEs will use beamforming, where a beamforming vector $\boldsymbol{w}_s \in \mathbb{C}^{(S \times 1)}$ must be designed for each satellite $s$. These vectors, along with the self-dependent RUE symbols $s_u \in \mathcal{C}\mathcal{N}(0,1)$ (i.e., assumed to be Gaussian distribution with zero mean and unit variance), are combined to form the transmitted vector $\boldsymbol{x}$, which is considered to be independent and satisfy  $\mathbb{E} [|x_u|^2] = 1$. For the considered  channel model, the received signal at RUE $u$ will be:
\begin{equation}
    y_u = \chi_{s,u} \boldsymbol{h}_{s,u} \boldsymbol{w}_{s,u} x_u + \sum_{s\neq s'} \sum_{u\neq u'} \chi_{s,u'} \boldsymbol{h}_{s,u} \boldsymbol{w}_{s,u'} x_u'  + n_u, \label{received_signal}
\end{equation}
where $h_{s,u}\in \mathbb{C}^{(S \times 1)}$ is the channel vector between satellite $s$ and RUE $u$, $\boldsymbol{w}_{s,u} \in \mathbb{C}^{(S \times 1)}$ represents the beamforming vector, and $x_u$ is the requested data of RUE $u$, which is assumed to be independent and satisfy $\mathbb{E} [|x_u|^2]=1$. The first term in (\ref{received_signal}) is the intended data for RUE $u$, the second term is the interference coming from the communication services for the other regions RUEs' links, and $n_u$ is the complex additive white Gaussian noise following the distribution $\mathcal{CN} (0, \sigma^2)$, where $\sigma^2$ is the noise power. Therefore the signal-to-interference-plus-noise ratio (SINR) at RUE u for each time slot depends on the specific $\boldsymbol{w}_s$ and the channel characteristics in (\ref{channel_model}) are defined as:
\begin{equation}
    \gamma_{s,u} = \frac{|\chi_{s,u}\boldsymbol{h}_{s,u} \boldsymbol{w}_{s,u}|^2}{\sigma^2 + |\sum_{s\neq s'} \sum_{u\neq u'} \chi_{s,u'} \boldsymbol{h}_{s,u'} \boldsymbol{w}_{s,u'}|^2}. \label{eq_sinr}
\end{equation}

We define a variable to optimally allocate the spectrum resource of each satellite $s$ according to the RUE $u$ QoS requirement. The spectrum allocation variable $\eta_{s,u}$, is defined such that $\eta_{s,u} \in \{0,1\}$ whether the satellite-RUE link is using the spectrum of satellite $s$ according to their QoS requirement and to maximize the system utility (to be described in an upcoming section). Each link is constrained to simultaneously select, for transmission according to the association constraint $\chi_{s,u}$. Each satellite spectrum constraint should follow this condition:
\begin{equation}
    \sum_{u\in\mathcal{U}} \chi_{s,u} \cdot \eta_{s,u}  \cdot B_{s,u}  \leq B^{\textrm{tot}}, \quad \forall s \in \mathcal{S}.
\end{equation}
The expression for the transmission rate of the satellite-RUE link data is as follows:
\begin{equation}
    R_{s,u} = \eta_{s,u}B^{\textrm{tot}} \log_2 (1+\gamma_{s,u}).
\end{equation}

\subsection{Spectrum Efficiency (SE) Model}
SE is used to evaluate the spectrum utilization of a multi-satellite system which is quantified as the sum of the transmission rates of all links divided by the total bandwidth spectrum consumption. The SE for each satellite and associated RUEs links can be expressed as:
\begin{equation}
    \Gamma_{s,u} = \frac{\sum\limits_{\substack{u\in\mathcal{U}}}R_{s,u}}{ 
    \sum\limits_{\substack{_{u\in\mathcal{U}}}} B_{s,u}}, ~\forall s \in \mathcal{S}.
\end{equation}
The total network SE is defined as the sum of the SE for all satellites available in the given network constellations:
\begin{equation}
    \Gamma^{\mathrm{tot}} = \sum_{s\in \mathcal{S}} \Gamma_{s,u}.
\end{equation}

In the decentralized resource allocation framework for satellites, the delay of satellite-RUE links is only considered in terms of transmission delay, excluding other types of scheduling delay at the medium access control (MAC) layer. Therefore, the delay constraints for satellite-RUE links are defined as:
\begin{equation}
    \sum_{u \in \mathcal{U} }  \chi_{s,u} R_{s,u} \geq \frac{D_{s,u}}{\tau_{s,u}}, \quad \forall s \in \mathcal{S},
\end{equation}
where $D_{s,u}$ denotes the remaining data to be transmitted from satellite $s$ to RUE $u$, and $\tau_{s,u}$ represents the residual latency, subject to the constraint $\tau_{s,u} \leq \tau_{\text{max}}$. In other words, $\tau_{s,u}$ cannot exceed the maximum delay allowed, $\tau_{\text{max}}$. For each satellite-RUE communication link, we can formulate the reliability constraints as follows:
\begin{equation}
     \gamma_{s,u} \geq  \gamma^{\mathrm{min}}, ~\forall u \in \mathcal{U},~\forall s \in \mathcal{S}.
\end{equation}
where $\gamma^{\textrm{min}}$ signifies the SINR minimum threshold specific to the satellite RUE receivers in the link.

\subsection{Problem Formulation}
\label{prob_form}
We next formally pose our problem. Specifically, our decentralized satellite management scheme focuses on optimizing beamforming and resource allocation for satellite-RUE links, while S2I links remain under centralized TGS control.
By carefully allocating satellite resources to each satellite-RUE link and accounting for the effects of S2I links, our goal is to enhance the SE, reduce co-channel interference, and improve the overall reliability of the satellite-RUE communication network. The optimization objective in this context is to maximize the network SE for the hybrid spectrum access system. This goal is achieved while considering limits on delay, reliability, maximum power, and spectrum allocation. Therefore, we will formulate a joint optimization problem whose goal is to maximize the total system SE of the multi-satellite cooperative network. We achieve this by jointly optimizing: the beamforming vector for each satellite-RUE link, denoted by $\boldsymbol{w}_{s,u}$, the satellite-to-RUE link selection, denoted by $\boldsymbol{\chi}_{s,u}$, and the spectrum allocation for each satellite-to-RUE link, denoted by $\boldsymbol{\eta}_{s,u}$. We formulate the objective function and associated constraints as follows:
\begin{maxi!}[2]<b>
{\substack{\{\boldsymbol{\chi}_{s,u}, \boldsymbol{w}_{s,u}, \boldsymbol{\eta}_{s,u}  \}  }}   
{ \Gamma^{\mathrm{tot}} \label{obj1} } {\label{opt:P1}} {\textbf{}}
\addConstraint{ \sum_{u \in \mathcal{U} }  \chi_{s,u} R_{s,u} \geq \frac{D_{s,u}}{\tau_{s,u}}, \quad \forall s \in \mathcal{S}  {\label{C1}}}
\addConstraint{  \gamma_{s,u} \geq  \gamma^{\mathrm{min}}, ~\forall s \in \mathcal{S}, \forall u \in \mathcal{U} {\label{C2}}}
\addConstraint{ \sum_{k \in \mathcal{K} } \chi_{s,u}||\boldsymbol{w}_{s,u}||^2 \leq P^{\mathrm{max}} {\label{C3}}}
\addConstraint{ \sum_{u \in \mathcal{U} }\chi_{s,u} = 1, ~\forall s \in \mathcal{S}  {\label{C4}}}
\addConstraint{  \chi_{s,u} \in \{0,1\}, ~\forall s \in \mathcal{S}, \forall u \in \mathcal{U} {\label{C5}}}
\addConstraint{\sum_{u}\chi_{s,u} \leq N^{\textrm{beam}}_s,~\forall s \in \mathcal{S},   {\label{C6} } }
\addConstraint{  \sum_{u\in\mathcal{U}} \chi_{s,u} \cdot \eta_{s,u}  \cdot B_{s,u}  \leq B^{\textrm{tot}}, \quad \forall s \in \mathcal{S}  {\label{C7} } }
\addConstraint{ \sum_{u\in\mathcal{U}} \chi_{s,u} \cdot R_{s,u}  \leq R^{\textrm{back}}, \quad \forall s \in \mathcal{S},   {\label{C8} } }
\end{maxi!}
where the objective function pertains to maximizing the network SE in (\ref{obj1}) while adhering to a number of critical constraints. Constraint (\ref{C1}) guarantees a minimum data rate for each satellite-RUE link based on the allowable delay. Constraint (\ref{C2}) enforces a minimum SINR threshold for each satellite-RUE link, ensuring reliable data transmission. Constraint (\ref{C3}) limits the maximum transmit power per satellite to $P^{\text{max}}$. Constraints (\ref{C4}) and (\ref{C5}) govern satellite-RUE link selection, ensuring each RUE connects to a single satellite and preventing multiple simultaneous connections. Constraint (\ref{C6}) restricts the number of connected RUEs per satellite not to exceed the maximum number of available spot beams $N^{\text{beam}}$. Constraint (\ref{C7}) ensures that the total utilized spectrum by a single satellite remains within the maximum available limit $B^{\text{tot}}$. Finally, constraint (\ref{C8}) enforces a dynamic backhaul capacity limitation on the association between each RUE u and its serving satellite. This constraint guarantees that the total data transmitted from all associated RUEs does not exceed the available backhaul capacity of the satellite. The backhaul capacity is considered to be dynamic, adapting based on the availability of access to the TGS. When a TGS connection is established, the backhaul capacity increases. Conversely, the capacity is reduced when the satellite relies solely on S2S links for backhaul communication.
\begin{figure}[t]
\centering
\captionsetup{justification=centering,singlelinecheck=false}
\includegraphics[width=0.8\columnwidth]{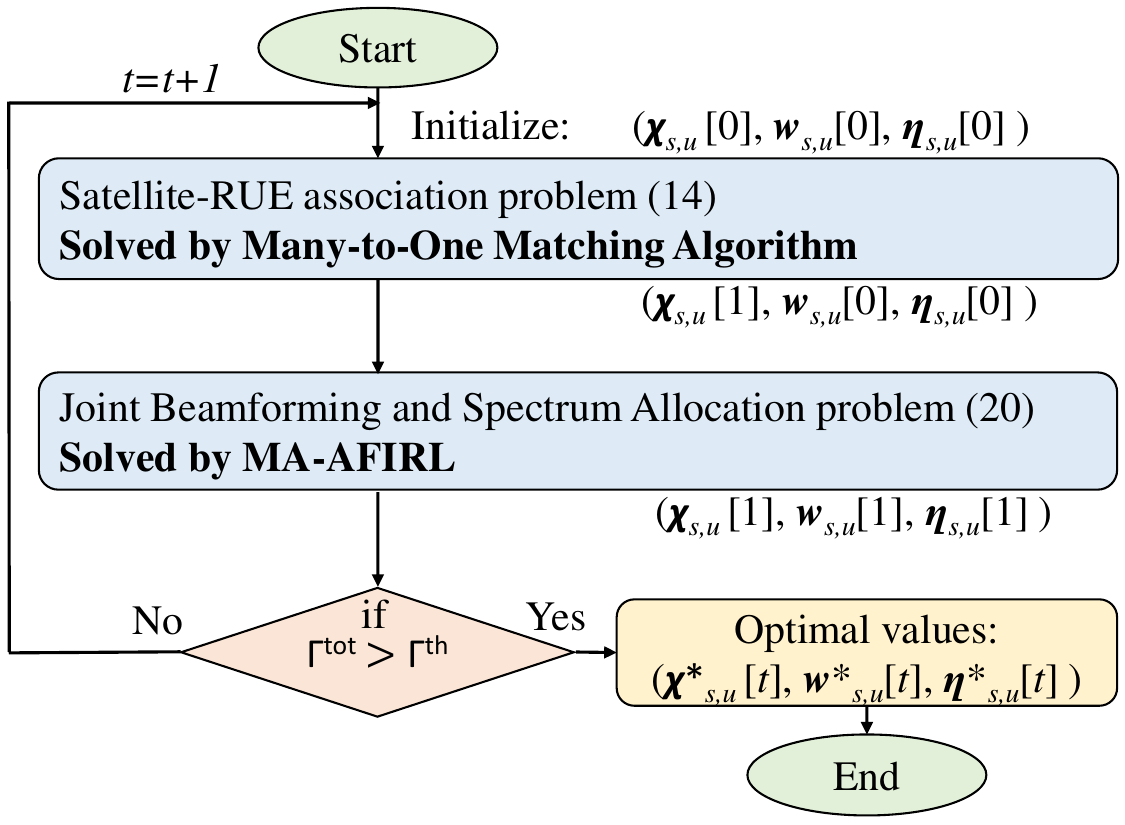}
\caption{Flowchart illustrating the proposed algorithms for solving joint beamforming and resource management.}
\label{flow_chart}
\end{figure}
\section{Joint Optimization of RUE Association, Beamforming and Spectrum Allocation}
\label{sol_app}
The formulated optimization problem in (\ref{opt:P1}) cannot be solved directly because the objective function in (\ref{obj1}) is non-convex and the beamforming vectors $\boldsymbol{w}_{s,u}$ are coupled with the link indicators $\boldsymbol{\chi}_{s,u}$ in the objective function and constraint (\ref{C3}). Therefore, we propose to solve it by the following steps:
\begin{enumerate}
\item \textbf{Many-to-one matching for satellite-RUE link association:} The algorithm begins by obtaining the channel realizations $\boldsymbol{h}_{s,u} [t,f]$. These realizations represent the time-frequency varying characteristics of the channels between each satellite-RUE link. Leveraging this information, we identify an initial feasible solution for the beamforming vectors, $\boldsymbol{w}_{s,u}$, and spectrum allocation vectors, $\boldsymbol{\eta}_{s,u}$.  With this initial solution in place, the remaining challenge lies in the binary link association problem, which determines which RUE connects to which satellite. To address this efficiently, we employ a many-to-one matching scheme based on game theory. This approach is particularly well-suited for handling binary decision problems like link association. 
\item \textbf{MA-AFIRL for joint beamforming and spectrum allocation:} Following the link association step, we have the optimal association variable, denoted by $\boldsymbol{\chi}_{s,u}^*$. This variable indicates which RUE is assigned to each satellite. With this information established, we utilize our novel MA-AFIRL algorithm to solve the remaining problem of jointly optimizing the beamforming vectors, $\boldsymbol{w}_{s,u}$, and spectrum allocation vectors, $\boldsymbol{\eta}_{s,u}$.  
\end{enumerate}
Many-to-one matching was chosen as the algorithm for associating satellites with RUEs due to its effectiveness in association management capabilities that ensure scalability and stability of connections by minimizing interference among users matched with satellites. Moreover, the MA-AFIRL method promotes efficient allocation of resources that can easily adapt without depending on humans as satellites improve their decision-making abilities over time. However, it's important to note that even though these algorithms are becoming more advanced and reaching a peak performance level, they still perform sub-optimally but they do outperform many of the traditional methods that were previously used in similar contexts. A detailed description of each proposed solution approach
is presented in the following subsections.

\subsection{Satellite-RUE Association Subproblem}
\label{sol_matching}
To address the computational complexity of problem (\ref{opt:P1}), we propose a decomposition approach. Before beamforming optimization, we focus on simplifying the problem by decoupling the satellite-RUE association.  Given initial beamforming and spectrum allocation vectors, we first tackle the satellite-RUE association problem, which can be reformulated as a matching problem:
\begin{maxi!}[4]<b>
{\substack{\{\boldsymbol{\chi}_{s,u}\}  }}   
{ \Gamma^{\mathrm{tot}} \label{obj1.1} } {\label{opt:P1.1}}{\textbf{}}
\addConstraint{ (\ref{C1})-(\ref{C6}),~\textrm{and},(\ref{C8}).  \label{C1.1}}
\end{maxi!}
To solve the problem in (\ref{opt:P1.1}), we propose a many-to-one matching for RUE association in subsequent sections.
\subsubsection{Many-to-One Matching based Satellite-RUE Association}
The RUE association problem can be modeled as a many-to-one matching game \cite{ren2021matching}. The concept of many-to-one matching refers to a scenario where agents belonging to one group can be matched with multiple agents from another group. In this game, there are two disjoint sets of players: the set of RUE clusters $k$ which is a subset of $\mathcal{U}$ (i.e., RUE clustering is performed based on proximity and in the same area of interest), and the set of satellites $\mathcal{S}$. 

In our proposed matching game, each RUE cluster $k \in \mathcal{U}$ can be associated with at most one satellite. Moreover, each satellite can serve a certain number of clusters depending on the maximum number of allowable clusters $Q^s$ at satellite $s \in \mathcal{S}$. Assume that each RUE cluster $k$ has a preference list $P(k)=s_{3}, s_{2}, s_{1}, s_{4}, ...$ which is a list in which satellites are sorted in order of RUE cluster $k$'s preference. For example, RUE cluster $k$ prefer satellite $s_{3}$ to satellite $s_{2}$. Therefore, many-to-one matching can be defined as:\\
\textbf{Definition 1.} \textit{A matching $\mu$ is a function from the set $\mathcal{S} \cup \mathcal{K}$ into the set of unordered families of elements of $\mathcal{S} \cup \mathcal{K}$ such that:\\
(1) $|\mu(k)|=1$ for each $k \in \mathcal{K}$ and $\mu(k)=k$ if $u$ is unassociated;\\
(2) $|\mu(s)| \leq Q^{s}$ for each satellite $s \in \mathcal{S}$;\\
(3) $\mu(k)=s$ if and only if $k$ is associated with $s$ and is an element of $\mu(s)$.}\\
where $|\mu(s)|=l$ means that $l$ RUE clusters are associated with satellite $s$ and $\mu(s)=\left\{ k_{1}, k_{2},...,k_{l} \right\}$ is a set of RUE clusters associated with satellite $s$.

For maximizing the achievable network SE of RUEs, each satellite $s$ determines the achievable network SE with each RUE cluster $k$ and has a preference list in high order. In the many-to-one matching problem of RUE association, we define the preference of each RUE cluster $k$ associated with satellite $s$ as follows:
\begin{equation}
U_{s,k} = \sum_{u \in C_{k}} \Gamma^{s}_{k,u}. \label{eq_pre_cluster}
\end{equation}
Then we define the preference of each satellite $s$ as:
\begin{equation}
U_{s} = \sum_{k \in \mathcal{K}}\sum_{u \in C_{k}} \Gamma^{s}_{k,u}. \label{eq_pre_ap}
\end{equation}
Therefore, in this matching, each satellite $s$ has a strict preference ordering $\succ_{s}$ over $\mathcal{K}$. 

Each satellite cluster also has a preference relation $\succ_{k}$ over the set $\mathcal{S} \cup \left\{0\right\}$, where $\left\{0\right\}$ denotes the RUE cluster is unmatched. Specifically, for a given RUE cluster $k$, any two satellites $s$ and $s'$ with $s, s' \in \mathcal{S}$, any two matching $\mu$ and $\mu'$ are defined as follows:
\begin{equation}
(s,\mu) \succ_k (s',\mu') \Leftrightarrow U_{s,k}(\mu) > U_{s,k}(\mu'),
\end{equation}
which indicates that the RUE cluster $s$ prefers satellite $s$ in $\mu$ to satellite $s'$ in $\mu'$ only if the RUE cluster $k$ can achieve a higher rate on satellite $s$ than satellite $s'$. Analogously, for any satellite $s$, its preference over the RUE cluster set can be described as follows. For any two subsets of RUE clusters $\mathcal{K}$ and $\mathcal{K}'$ with $\mathcal{K} \neq \mathcal{K}'$, any two matching $\mu$ and $\mu'$ with $\mathcal{K} = \mu(s)$ and $\mathcal{K}' = \mu'(s)$ are defined as:
\begin{equation}
(\mathcal{K},\mu) \succ_s (\mathcal{K}',\mu') \Leftrightarrow U_{s}(\mu) > U_{s}(\mu'),
\end{equation}
which represent that satellite $s$ prefers the set of RUE clusters $\mathcal{K}$ to $\mathcal{K}'$ only when satellite $s$ can get a higher rate from $\mathcal{K}$. 

Due to the existence of peer effects and non-substitutability in (\ref{eq_sinr}), the preference lists of players change frequently during the matching process, which makes it difficult to design the matching processes. To handle the peer effects and ensure exchange stability, given a matching function $\mu$, and assume that $\mu(i)=n$ and $\mu(j)=m$, we define the swap matching as:
\begin{equation}
\mu^{j}_{i}=\left\{ \mu \setminus \left\{(i,n),(j,m)\right\} \bigcup \left\{(j,n),(i,m)\right\} \right\},
\label{eq_swap_matching}
\end{equation}
where RUE clusters $i$ and $j$ exchange their matched satellites $n$ and $m$ while keeping all other matching states the same. Based on the swap operation in (\ref{eq_swap_matching}), we define the concept of swap-blocking pair as follows \cite{cui2017optimal}.\\
\textbf{Definition 2.} \textit{A matching $\mu$ is two-sided exchange-stable if and only if there does not exist a pair of RUE clusters $(i,j)$ such that:\\
1) $\forall k \in \left\{i,j,n,m\right\},U_{k}(\mu^{j}_{i}) \geq U_{k}(\mu)$;\\
2) $\exists k \in \left\{i,j,n,m\right\},$ such that $U_{k}(\mu^{j}_{i}) > U_{k}(\mu)$,}\\
where $U_{k}(\mu)$ represents the utility of RUE cluster $k$ under matching $\mu$.
However, swap matching depends on the performance given the initial matching. Therefore, to achieve the goal of stable initial matching, the deferred acceptance (DA) algorithm was proposed and applied to the marriage markets and college admission problems \cite{gale1962college}. 

In the DA algorithm, the agents on one group propose a pair formation with the agents of the other side group according to their preference, and an iterative procedure. At first, the matching procedure starts with building preference lists, i.e., $P(k)$ for RUE cluster $k$. In each iteration, each RUE cluster $k$ proposes to highest preferred satellite for association in the preference list. Each satellite $s$ accepts the proposal of RUE cluster $k^{*}$ with the highest rank among the RUE cluster's proposal list $G(s)$. At the same time, exclude satellite $s$ from RUE cluster $k^{*}$'s preferred list. However, when the current number of proposals for satellite $s$ is greater than the maximum capacity, i.e., $|\mu(s)| \geq Q^{s}$, if a proposal from RUE $k^{*}$ comes, no further proposals are accepted and satellite $s$ is excluded from the preference list of RUE cluster $k^{*}$. The iterative approach continues until every RUE cluster has gotten an acceptable proposal, at which point a stable solution to the RUE association problem is attained. The stability and convergence in many-to-one matching algorithms are not always guaranteed but can be achieved under certain conditions. While stability is possible if the algorithm aims for a stable matching, dynamic factors such as changing user demands or satellite capacities may disrupt it, causing the system to become unstable over time.

Finally, the output of the many-to-one matching $\mu$ is transformed to the RUE association vector $\boldsymbol{\chi}^*$. The complexity of a many-to-one matching game-based DA algorithm depends on the required number of accepting/rejecting decisions to attain the stable matching $\mu$. Each RUE cluster in the network recommends an association with the satellite at the top of their preference list during each iteration. The satellite then decides whether to accept or reject the proposal. Each satellite cluster's preference list can only be as large as $|\mathcal{S}|$ in this case. As a result, the stable matching converges in $\mathcal{O}(|\mathcal{K} \times \mathcal{S}|)$ iterations, where $\mathcal{K}$ and $ \mathcal{S}$ are the number of RUEs and satellites in the considered network. Algorithm \ref{alg:association} describes the many-to-one matching based on the DA algorithm for RUE association.

\begin{algorithm}[t]
	\caption{\strut Many-to-One Matching for Satellite-RUE Association} 
	\label{alg:association}
	\begin{algorithmic}[1]
	    \STATE{\textbf{Step-I:} Initial matching based on DA algorithm}
        \STATE{The sorted preference lists $P(k)$ for all RUE clusters based on the scalar channel gain $\sum_{u \in C_{k}}|\hat{h}_{s,u}|^{2}$.}
        \STATE{$\mathcal{K}_{\textrm{cur}} \leftarrow \mathcal{K}$}
        \WHILE{$\mathcal{K}_{\textrm{cur}} \neq \emptyset$}
        \STATE{$G(s) \leftarrow \emptyset, \forall s \in \mathcal{S}$}
        \FOR{$k \in \mathcal{K}_{\textrm{cur}}$}
        \STATE{$G(s)=G(s) \cup \left\{k\right\}$, with $P(k)[0]=s$}
        \ENDFOR        
        \FOR{$s \in \mathcal{S}$}
        \IF{$|G(s)| \neq 0$}
        \STATE{Sort $G(s)$ by the preference with the scalar channel gain $\sum_{u \in C_{k}}|\hat{h}_{s,u}|^{2}$}
        \STATE{$k^{*} = G(s)[0]$}
        \IF{$|\mu(s)|<Q^{s}$}
        \STATE{$\mu(s)=\mu(b)\cup \left\{k^{*}\right\}$}
        \STATE{$\mu(k^{*})=s$}
        \STATE{$\mathcal{K}_{\textrm{cur}} = \mathcal{K}_{\textrm{cur}} \setminus \left\{k^{*}\right\}$}
        \ELSE
        \STATE{$P(k^{*})=P(k^{*}) \setminus \left\{s\right\} $}
        \ENDIF
        \ENDIF
        \ENDFOR        
        \ENDWHILE
		\STATE{\textbf{Output:} The initial matching function $\mu^{0}$.}
	    \STATE{\textbf{Step-II:} Two-sided exchange stable matching for Satellite-RUE Association}
        \STATE{The initial matching function $\mu^{0}$.}
        \REPEAT
        \STATE{$\forall$ RUE cluster $k \in \mu^{0}$, it searches for another RUE cluster $k' \in \mu^{0} \setminus \mu^{0}(\mu(k))$ to check whether $(k,k')$ is a swap-blocking pair.}
        \IF{$(k,k')$ is a swap-blocking pair}
        \STATE{Update $\mu = \mu^{k'}_{k}$}
        \ELSE
        \STATE{Keep the current matching state}
        \ENDIF
        \UNTIL{No swap-blocking pair can be constructed.}
\STATE{\textbf{Output:} The optimal matching function $\mu^{*} \rightarrow$ RUE association vector $\boldsymbol{\chi}^*$.}
	\end{algorithmic}
\end{algorithm}

\subsection{Joint Beamforming and Spectrum Allocation Subproblem}
After obtaining optimal satellite-RUE association $\boldsymbol{\chi}^*$, we solve beamforming and spectrum allocation optimization. Given optimal $\boldsymbol{\chi}^*$ vectors, the remaining problem can be reformulated as follows:
\begin{maxi!}[4]<b>
{\substack{\{\boldsymbol{w}_{s,u}, \boldsymbol{\eta}_{s,u}  \}  }}  
{ \Gamma^{\mathrm{tot}} \label{obj1.2} } {\label{opt:P1.2}}{\textbf{}}
\addConstraint{ (\ref{C1}), (\ref{C2}), (\ref{C3}),(\ref{C7}),~\textrm{and},~(\ref{C8}}).  \label{C1.2}
\end{maxi!}
To solve the subproblem in (\ref{opt:P1.2}), we propose an MA-AFIRL algorithm for beamforming and spectrum allocation in a subsequent section.
\begin{figure*}[t]
        \centering
        \captionsetup{justification=centering,singlelinecheck=false}
        \includegraphics[width=0.8\linewidth, height=4in]{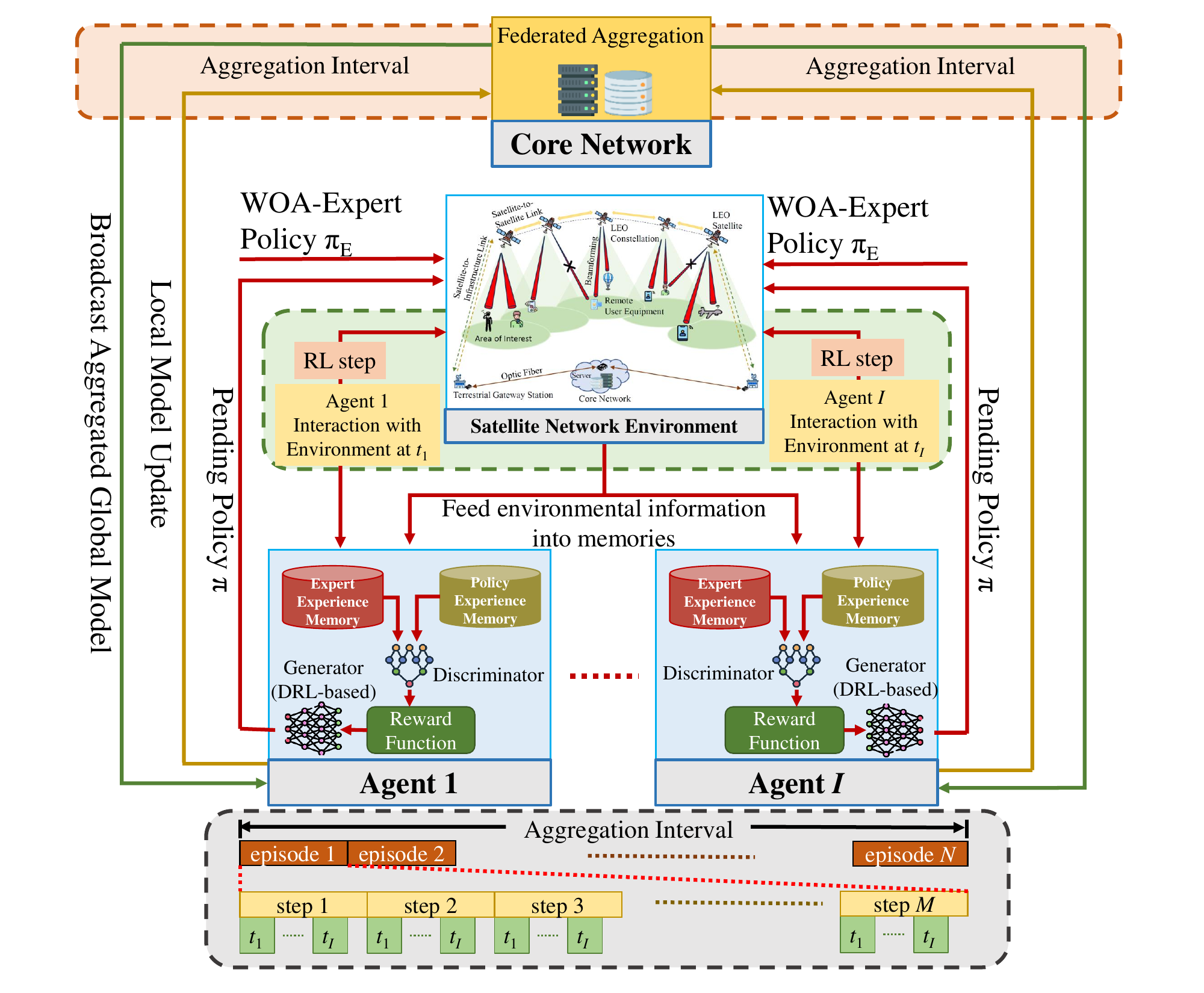}
        \caption{Proposed multi-agent asynchronous federated inverse reinforcement learning (MA-AFIRL) learning framework.}
        \label{framework}
    \end{figure*}
\subsubsection{Proposed MA-AFIRL Algorithm's Rationality}
\label{sol_IRL}
We present a novel distributed multi-satellite resource optimization algorithm based on MA-AFIRL. The algorithm frames the optimization problem as a MDP and employs a GAIL architecture. Unlike centralized approaches, MA-AFIRL eliminates the need for satellites to upload raw data to a central server, reducing transmission overhead and delay. The proposed MA-AFIRL framework operates within a client-server (i.e., satellite-TGS) architecture, as depicted in Fig.~\ref{framework}. Satellite agents act as clients, while the TGS serves as the central server. The method implements a client-server architecture that combines IRL and FL, using a SE maximization strategy for multi-satellite systems. Aggregating average algorithms create a global model with strong generalization capabilities. MA-AFIRL enables efficient, adaptive resource allocation without human intervention as satellites improve their decision-making abilities. While this approach outperforms many traditional methods, it still operates sub-optimally despite its sophistication. Subsequent sections provide detailed descriptions of the proposed solution approaches.

Each satellite makes independent decisions based on its local observations, without knowledge of other satellite's actions. This limited information can lead to conflicts over resource allocation and performance degradation if all satellites make decisions simultaneously. We introduce an asynchronous approach that divides the training process into $t=I$ discrete time slots. Each satellite is assigned a specific time slot $t$ during which it independently performs CSI sensing and takes actions. This ensures that each satellite has dedicated time for decision-making without interference from others, effectively minimizing resource conflicts. It is important to note that the data transmission of each satellite continues throughout an episode, until successful transmission or episode completion. The designated time slot $t_i$ only applies to beamforming and spectrum allocation, not data transmission itself. This separation allows for efficient resource allocation during satellite-RUE communication.

\subsubsection{IRL Learning Architecture}
The IRL framework utilizes ANN to recover the reward function implicitly optimized by an expert policy. Given an expert policy demonstrated by a skilled agent, the IRL approach assigns higher reward values to actions that the expert takes and lower values to alternative actions. Moreover, traditional approaches to joint beamforming and spectrum allocation optimization face challenges when dealing with non-convex problems. MA-RL and existing numerical optimization methods often struggle to find the global optimum in such scenarios, and may only achieve partial local optima after converting the non-convex problem to a convex form. While RL can be effective in finding optimal solutions for less complex communication problems, the convergence speed and even guarantee of convergence become unreliable as the complexity of the agent's observations, actions, and environment increases. Thus this work proposes an IRL approach to address these limitations and the rationale is given as follows:
\begin{itemize}
    \item Instead of explicitly searching for the global optimum, IRL focuses on learning a reward function that encourages the learning agent to reach desirable states. This simplifies the problem and avoids the complexities associated with directly finding the global optimum in non-convex problems.
    \item Utilizing an expert policy involves using a numerical optimization technique, such as the WOA, as an \emph{expert policy} agent. The IRL agent learns by mimicking the expert's behavior while aiming to improve upon it.
    \item As the IRL agent progresses in its learning, the expert's influence is gradually reduced, allowing the agent to explore beyond the expert's solutions and potentially discover the global optimum.
\end{itemize}
This combination of IRL and an expert policy offers a promising alternative for tackling the complexities of joint beamforming and spectrum allocation optimization in non-convex environments. This approach essentially encodes the WOA expert's knowledge and preferences into the reward function, enabling the learning agent to mimic the expert's behavior. This expert-based policy can be defined as follows:
\begin{equation}
    \underset{{r}}{\mathrm{max}} ~ \Big( \mathbb{E}_{\pi_{E}} \big[r(s, a)\big] - 
    \Big(   \underset{{r}}{\mathrm{max}}~\mathbb{E}_{\pi} \big[r(s, a)\big] + H(\pi)
    \Big)\Big).
     \label{policy}
\end{equation}
Let $\pi_e$ represent the expert policy, and $\pi$ denote the policy being learned. The expected discounted reward under the expert policy is denoted as $\mathbb{E}_{\pi_e} \left[ \sum_{t=0}^{\infty} \gamma^t r(\mathbf{s}_t, a_t) \right]$, where $\gamma$ is the discount factor and $r(\mathbf{s}_t, a_t)$ is the reward received at state $\mathbf{s}_t$ after taking action $a_t$. The causal entropy of the policy being learned is represented by $H(\pi) = \mathbb{E}_{\pi_e} \left[ -\log \pi (a|\mathbf{s}) \right]$.

To enhance the training efficiency of the IRL framework, we adopt the GAIL framework \cite{ho2016generative} as our IRL model. GAIL leverages the strengths of GANs to learn the policy from expert demonstrations directly. The GAIL-based IRL approach comprises a discriminator and a generator network. The discriminator network acts as an \emph{expert} aims to discern whether a state-action pair originates from the expert policy or the generated policy, thereby implicitly defining the reward function. In contrast, the generator network attempts to generate policy outputs indistinguishable from the expert's actions. To optimize the discriminator, we rewrite the objective function in (\ref{policy}) as:
\begin{equation}
    \underset{{\mathcal{D}}}{\mathrm{max}} ~ \Big( \mathbb{E}_{\pi_{E}} \big[\log \big( 1-\mathcal{D}(s, a) \big)\big] + 
    \Big(   \underset{{\pi}}{\mathrm{min}}~\mathbb{E}_{\pi} \big[\log \mathcal{D}(s, a)\big] + H(\pi)
    \Big)\Big),
     \label{discr_reward}
\end{equation}
where $\mathcal{D}$ represents the discriminator network, tasked with differentiating between expert actions generated by $\pi_e$ and actions generated by the policy being learned, $\pi$. The discriminator output $\mathcal{D}(\mathbf{s}, a)$ lies between 0 and 1. Ideally, the discriminator should output close to 0 for expert actions and close to 1 for non-expert actions. To incentivize the policy to mimic the expert, we leverage the discriminator's output to define a reward function:
\begin{equation}
    r(s, a) = - \log \mathcal{D}(s, a).  \label{reward}
\end{equation}

Consequently, the parameters of the discriminator network $\theta$ are updated using gradient descent, according to the following equation.
\begin{equation}
    \mathcal{L}_{\mathrm{dis}}(\theta) = \mathbb{\hat{E}}_{\tau_{\pi}} \big[ \nabla_{\boldsymbol{\theta}} \log \mathcal{D}_{\boldsymbol{\theta}}(s, a)  \big] + \mathbb{\hat{E}}_{\tau_{\pi}} \big[ \nabla_{\boldsymbol{\theta}} \log \big( 1 - \mathcal{D}_{\boldsymbol{\theta}}(s, a)  \big) \big]. \label{discr_gradient}
\end{equation}
The generator network continuously seeks to improve the pending policy by approximating the expert's behavior. The generator network parameters, denoted by $\phi$, are updated using gradient descent via the following equation:
\begin{equation}
     \mathcal{L}_{\mathrm{gen}}(\phi) = \mathbb{\hat{E}}_{\tau_{\pi}} \big[ \nabla_{\boldsymbol{\phi}} \log \pi_{\boldsymbol{\phi}}(a|s) Q(s, a)  \big] - \nabla_{\boldsymbol{\phi}} H (\pi_{\phi}), \label{gener_gradient}
\end{equation}
where $Q(s, a) = \mathbb{\hat{E}}_{\tau_{\pi}} \big[  \log \big( \mathcal{D}(s, a)| s_0 = s, a_0 = a \big)  \big]$.
To further improve understanding of the IRL framework, a diagram is presented in Fig. \ref{framework}. During the training stage, the discriminator network plays two crucial roles: (1) differentiating between expert actions and actions generated by the policy being learned and (2) utilizing this distinction to generate rewards that guide the learning process. Meanwhile, the generator network continuously seeks to improve the pending policy by approximating the expert's behavior. The generator iteratively generates policies and receives feedback in the form of rewards from the discriminator, allowing it to gradually refine its actions until they closely resemble those of the expert.

\subsubsection{Design of MDP}
Analogous to the RL-based approaches, the problem in (\ref{opt:P1}) requires reformulation as a MDP. This requires defining a $4$-tuple $(\mathcal{S}, \mathcal{A}, r, \gamma)$ to represent the MDP, where:
\begin{itemize}
    \item $\mathcal{S}$: The state space encompasses all possible states that the agent can encounter.
    \item $\mathcal{A}$: The action space, represents all available actions the agent can take.
    \item $r$: The dynamic reward function, assigning a numerical value to each state-action pair obtained by the novel GAIL framework.
    \item $\gamma \in [0, 1)$: The discount factor, which determines the present value of future rewards.
\end{itemize}
This reformulation enables us to leverage a powerful IRL algorithm for solving (\ref{opt:P1}). At each time step $t$, the satellite acts as individual \emph{agents}, obtaining their current state from the state space based on local observations. The satellite then learns and updates its policy, represented by the action value function $Q(s_t, a_t)$, which guides its action selection. Subsequently, each satellite independently chooses its transmission beamforming and spectrum allocation. As a result of these collective actions, the environment transitions to a new state $s_{t+1}$, and each satellite receives a reward $r_t(i)$.

Due to the distributed nature of the multi-satellite and their limited local observations, horizontal federated learning is used to facilitate collaborative training in all agents \cite{FL}. During training, each satellite maintains its own GAIL framework. These networks are trained using small batch gradient descent, leveraging local observations and environmental interactions. Every $N$ episode, which consists of $M$ steps with each satellite sequentially interacting with the environment within their assigned time slot $t$, the model parameters are aggregated across all agents using the federated averaging (FedAvg) algorithm.

\indent \textbf{Network State Space:}
In particular, the state space of the network, denoted as $\mathcal{S}$, encompasses localized observational data relevant to resource allocation for each satellite, as indicated in \cite{ML-DRL_power}. The instantaneous local CSI for the satellite-RUE link is formally defined as follows:
\begin{equation}
    H_{s,u} =  \{h_{s,u}, h^{'}_{s',u'}  \}  
\end{equation}
where $h_{s,u}$ and $h^{'}_{s',u'}$ represent the channel gain of the main and interference satellite-RUE link, respectively. This state space of the network for satellite $s$ can be represented as:
\begin{equation}
     \mathcal{S}^t_s = \big\{ H_{s,,u}, \boldsymbol{\chi}^{*}_{s,u}, D_{s,u}, \tau_{s,u}, \boldsymbol{w}_{s,u}(t-1), \boldsymbol{\eta}_{s,u}(t-1)    \big\}_{u\in \mathcal{U}}. 
\end{equation}
where $\boldsymbol{w}_{s,u}(t-1)$ and $\boldsymbol{\eta}_{s,u}(t-1)$ represents the beamforming and spectrum allocation in the previous time step.\\
\indent \textbf{Network Action Space:}
The beamforming and spectrum allocation are inherently continuous which is most suitable and efficient for the IRL approach. Consequently, the action space A is defined as:
\begin{equation}
    \mathcal{A}^t_j = \big\{ \boldsymbol{w}_{s,u}(t), \boldsymbol{\eta}_{s,u}(t)  \big\}_{u\in \mathcal{U}} 
\end{equation}
where $\boldsymbol{w}_{s,u}(t)$ and $\boldsymbol{\eta}_{s,u}(t)$ represents the beamforming and spectrum allocation vector in each time step for all the satellite-RUEs links.

\indent \textbf{Network Reward Function:}
Our goal is to optimize the network SE while respecting constraints as mentioned in problem (\ref{opt:P1}). Unlike traditional approaches that rely on hand-crafted reward functions, the proposed IRL method leverages the WOA expert policy for reward generation. Recognizing the potential of the WOA approach to achieve local optima and even approximate global solutions, we adopt it as the expert policy. The rewards are then obtained from the discriminator network based on equation (\ref{reward}). For greater clarity, the proposed IRL approach is presented in Algorithm \ref{alg:AF-MAIDRL_learning}. A key aspect of IRL methods is the collection and storage of pending policy trajectories (training data) in an experienced pool. Once sufficient data has accumulated, the discriminator network is updated first, followed by the generator network using rewards generated by the discriminator. This iterative process enables the IRL approach to find a sub-optimal solution to the problem (\ref{opt:P1}) efficiently. 
\subsubsection{Federated Aggregation for Generalizability}
For efficient model training across geographically distributed agents, the proposed framework utilizes the federated averaging (FedAvg) algorithm to aggregate and distribute the satellite learning model updates. This approach allows each satellite to perform GAIL-based learning (with stochastic gradient descent (SGD) on its local data), while the core network server aggregates these updates and refreshes its parameters through a weighted averaging process. Here is how the core network parameters are updated:
\begin{enumerate}
    \item Collect model updates: Every N episode, each agent uploads its locally trained GAIL-based learned policy model parameters to the core network server.
    \item Weighted averaging: The core network server calculates a weighted average of the model parameters received, where the weights typically reflect the size of each agent's local data set or its contribution to the overall training process.
    \item Update core network server parameters: The core network server updates its parameters using the calculated weighted average.
\end{enumerate}

This iterative process enables the learned policy model to gradually improve its performance across all agents, leveraging the collective learning power of the distributed network. The following expression updates the global learned policy model parameters:
\begin{equation}
    \omega_{t+1} \leftarrow  \omega_{t} - \beta \underset{s \in \mathcal{S}}{\sum} \frac{M_s}{M} \nabla f_s (\omega_{t}) \label{fed_aver_v1}
\end{equation}
The core network server updates its neural network parameters, denoted by $\omega_{t}$, using the FedAvg algorithm with a learning rate of $\beta$. Each satellite $s$ contributes its local model updates based on a training batch size of $M_s$, forming a total batch size of $M$ across all agents. The gradient $ \nabla f_s (\omega_{t})$ is obtained from satellite $s$. Considering the specific characteristics of our satellite resource allocation problem, where the number of satellites is relatively small and communication requests are periodic, we maximize the aggregation effect by using an equal weight for each agent in the FedAvg update. This choice is motivated by two factors:
\begin{enumerate}
    \item Limited number of satellite agents: Since satellites prioritize direct communication with the TGS, S2S links are used sparingly. Therefore, assigning equal weights avoids biasing the update towards specific agents with potentially less relevant data.
    \item Periodic communication requests: All S2S links contribute their updates regularly, ensuring that the aggregated model incorporates the knowledge from the entire network.
\end{enumerate}
Therefore, we rewrite Equation (\ref{fed_aver_v1}) as follows, representing the MA-AFIRL aggregation:
\begin{equation}
    \omega_{t+1} \leftarrow  \omega^k_{t} - \beta  \nabla f_k (\omega^k_{t}), ~\forall k \in \mathcal{K}, \label{fed_aver_v2}
\end{equation}
\begin{equation}
    \omega_{t+1} \leftarrow \underset{k \in \mathcal{K}}{\sum} \frac{M_k}{M} \omega^k_{t+1},  \label{omega_update}
\end{equation}
where $\omega^k_{t}$ denotes the agent $k$ neural network parameters at time slot $t$.

\subsection{MA-AFIRL Learning and Online Implementation}
The MA-AFIRL framework operates in two distinct phases: the offline learning phase and the online execution phase. The computationally intensive learning phase, where the learned policy model is trained, takes place offline. Conversely, the execution phase, where the trained model is used for resource allocation, occurs online. The offline learning phase is made up of the following steps:
\begin{enumerate}
    \item Model Broadcast: The core network server transmits the initial or aggregated network model parameters to all satellite agents via S2S links.
    \item Distributed Exploration and Interaction: Each S2S link independently observes its local state, interacts with the environment using the DDPG algorithm (actor and critic networks, and experience replay), and gathers experiences.
    \item Distributed Learning: All satellite agents perform mini-batch stochastic gradient descent on their local data, then update their model parameters, and upload them to the core network server.
    \item Model Aggregation: The core network server receives the uploaded models, aggregates them based on the satellites' contributions (e.g., using FedAvg), and broadcasts the new model parameters to all satellite links.
    \item Convergence Check: The training process iterates through steps 1-4 until the network converges.
\end{enumerate}
However, the online execution phase consists of the following steps:
\begin{enumerate}
    \item Action Selection: Each satellite agent uses the trained GAIL-based model and its learned policy to select the optimal beamforming and spectrum allocation for data transmission.
\end{enumerate}
The specific steps of both the learning and execution phases are detailed in Algorithms \ref{alg:AF-MAIDRL_learning} and \ref{alg:AF-MAIDRL_implementation}, respectively.

\begin{algorithm}[t]
\caption{\strut Learning Process for MA-AFIRL Algorithm} 
\label{alg:AF-MAIDRL_learning}
\begin{algorithmic}[1]
\STATE{\textbf{Initialize:} start satellite network environment simulation, select satellites, initialize update interval, aggregate interval, and other parameters, initialize network parameters $\omega^t_k$ and target network parameters $\omega^t_{k'}$ randomly, policy experience memory $X_p$, discriminator network with parameter $\chi_0$, and generator network with parameter $\omega_0$}
\STATE{\textbf{Input:} Expert experience memory $X_E$, the total number of training episodes $N$, the maximum number of time slots $T$, and the update cycle $T_\mathrm{upd}$}
    \FOR{each learning episode $e = 1, 2, \cdots, N$}
    \STATE{Initialize state $s^0 \in \mathcal{S}$ with satellites' position and channel fading.}
    \STATE{Update $D_j = D$, $\tau_j = \tau$}
        \FOR{each time step $= 1, 2, \cdots, M$}
            \FOR{each agent $k$}
		\STATE{Select the action $\mathcal{A}^t_j$ by generator network;}
		\STATE{Observe the new state $\mathcal{S}_{t+1}$ in next step $t + 1$;}
		\STATE{Store ($\mathcal{S}_{t}, \mathcal{A}^t_j, \mathcal{S}_{t+1}$) into policy experience pool}
		\STATE{$\mathcal{S}_{t} \leftarrow \mathcal{S}_{t+1}$}
		\ENDFOR
        \ENDFOR
        \IF{$\frac{\mathrm{time~step}}{T_\mathrm{upd}} = 0$}
		\STATE{Sample random mini-batch of trajectory $\tau_{\pi_E}$ from $X_E$}
		\STATE{Sample random mini-batch of trajectory $\tau_{\pi_E}$ from $X_P$}
		\STATE{Update the discriminator parameter by (\ref{discr_gradient})}
        \STATE{Calculate the rewards with (\ref{reward})}
        \STATE{Update the generator parameter by (\ref{gener_gradient})}
		\ENDIF
        \IF{$\frac{\mathrm{time~step}}{\mathrm{aggregation~interval}} = 0$}
		\STATE{all the satellites update local models to the core network server}
		\STATE{core network server aggregates models following (\ref{fed_aver_v2}) (\ref{omega_update})}
        \STATE{core network server aggregates distributes the global model to satellites}
		\ENDIF
        \ENDFOR
		\STATE{\textbf{Output:} Optimal networks' parameters $\boldsymbol{\omega}_k$}
	\end{algorithmic}
\end{algorithm}
\begin{algorithm}[t]
	\caption{\strut Implementation Process for MA-AFIRL Algorithm} 
	\label{alg:AF-MAIDRL_implementation}
	\begin{algorithmic}[1]
	    \STATE{\textbf{Initialize:} satellite simulation environment and satellite numbers; update simulation parameters and load learned network's parameters $\boldsymbol{\omega}_k$ .}
		\FOR{each implementation episode $e=1,2, \cdots, N$}
		\STATE{set satellites' position and channel fading.}
        \STATE{Update $D_j = D$, $\tau_j = \tau$}
		\FOR{each time step $= 1, 2, \cdots, M$}
        \FOR{each agent $k$}
		\STATE{obtain state $\mathcal{S}_t(k)$, select beamforming $\boldsymbol{\omega_{s,u}(t)}$ and spectrum allocation vector $\boldsymbol{\eta}_{s,u}(t)$ at times slot $t$ based on generator action $\mathcal{A}_t(k)$}
		\STATE{update interference of S2I and S2S links}
		\ENDFOR
        \STATE{update channel fading}
        \ENDFOR
        \ENDFOR
		\STATE{\textbf{Output:} Optimal network SE.}
	\end{algorithmic}
\end{algorithm}

\section{Simulation Results and Analysis}
\label{simul}
In this section, we will carry out multiple simulations to illustrate the impact of IRL-GAIL in proposed satellite networks. First, we will outline the simulation settings. Then, we will compare IRL with competing methods using state-of-the-art (SOTA) techniques under various network conditions, thus demonstrating the effectiveness of our proposed adaptation and correlation strategies.

\subsection{Simulation Setup}
For our simulations, we deploy the RUEs across a $500 \times 500~$ square kilometer (km$^2$) area, serviced by a Walker satellite constellation comprising three orbital planes. A free-space channel model is employed for the communication environment. Moreover, the testing network topology is composed of $10$ satellite and $100$ RUEs. The satellites operate at an altitude of $500~$km with a speed of $7.5622~$km/s, while the RUEs move at a speed of $3~$km/h. The satellite propagation delay ranges from $2$ to $20~$ms depending on the satellite's position in orbit. Additionally, it is assumed that each RUE has different data requirements. To evaluate the performance of our proposed IRL-GAIL, we conducted extensive simulations using the Stable Baselines $3$ RL framework. This popular library provided a robust and flexible platform for training and testing various RL agents. By leveraging the pre-implemented algorithms and customizable environments within Stable Baselines $3$, we were able to efficiently explore different parameter configurations and compare our approach against established baselines. The key system and learning parameters used in the simulation are summarized in Table \ref{sim_tab} and Table \ref{learing_tab}, respectively.
\begin{table}[t]
\centering
\caption{Simulation Parameters}
\label{sim_tab}
\renewcommand{\arraystretch}{1} 
\begin{tabular}{|c|c|c|}
\hline
\textbf{Notation} & \textbf{Definition} & \textbf{Value} \\
\hline \hline
$f_c$ & Ka-band carrier frequency & 20 GHz \\ \hline
$B_u$ & Bandwidth & 500 MHz \\ \hline
$G_s$ & Satellite antenna gain & 33.13 dBi \\ \hline
$d_{s,u}$ & Satellite altitude & 500 km \\ \hline
$N_0$ & Noise power & -43 dB \\ \hline
$G_{u}$ & RUE antenna gain & 34.2 dBi \\ \hline
$p_u$ & Satellite transmit power & 10 W \\ \hline
$T_{\textrm{orb}}$ & Orbital period & 100 min \\ \hline
$\nu_{s,u}$ & Doppler shift & 20 kHz \\ \hline
$\textrm{SF}$ & Shadow fading & $\mathcal{C}\mathcal{N}(0,1)$ \\ \hline
$\textrm{CL}$ & Clutter loss & 0 dB \\ \hline
$\textrm{PL}_{g}$ & Atmospheric gas attenuation & -10 dB \\ \hline
$\textrm{PL}_{s}$ & Scintillation attenuation & -20 dB \\ \hline
\end{tabular}
\end{table}

\begin{table}[t]
\centering
\caption{Training Parameters}
\label{learing_tab}
\setlength{\tabcolsep}{3pt}
\renewcommand{\arraystretch}{1} 
\begin{tabular}{|c|c|}
\hline
\textbf{Parameter} & \textbf{Value} \\
\hline \hline
\multicolumn{2}{|c|}{RL - PPO Network} \\ \hline
Network architecture & $[128, 128]$ \\ \hline
Epochs & $64$ \\ \hline
Expert data size & $50000$ \\ \hline
\multicolumn{2}{|c|}{IRL - GAIL Network} \\ \hline
$\mathrm{Demo\_batch\_size}$\footnote{The number of samples in each batch of expert data. The discriminator batch size is twice this number because each discriminator batch contains a generator sample for every expert sample.} & $1024$ \\ \hline
$\mathrm{Gen\_replay\_buffer\_capacity}$\footnote{The capacity of the generator replay buffer (i.e., the number of obs-action-obs samples from the generator that can be stored). By default, this is equal to $gen\_train\_timesteps$, meaning that we sample only from the most recent batch of generator samples.} & $512$ \\ \hline
\multicolumn{2}{|c|}{Remaining parameters for all networks} \\ \hline
Activation function & ReLU \\ \hline
Total Step & $50000$ \\ \hline
\end{tabular}
\end{table}

\subsection{Analysis of Benchmarks}
The proposed \textit{IRL-GAIL} resource allocation algorithm is compared with the following three benchmarks, namely \textit{Expert}, \textit{RL-PPO}, and \textit{Fairness}.
\begin{itemize}
    \item \textit{IRL-GAIL}: The proposed algorithm employs federated learning within a multi-agent satellite system for cooperative training. During inference (i.e., testing), it leverages GAIL-based IRL, where each satellite agent makes decisions based on its learned policy. For consistency in comparison, the network architecture used here is identical to the one employed in the \textit{RL-PPO} algorithm, enabling a fair evaluation within the same network environment.
    \item \textit{RL-PPO}: This algorithm integrates federated learning with the multi-agent PPO framework. In this scenario, each agent performs traditional RL using PPO, with federated learning facilitating the training of the neural networks, which are then used for inference.
    \item \textit{Expert}: This benchmark employs the WOA to generate a set of $50000$ demonstrations, representing the optimal decision-making strategy for the given network environment. These demonstrations serve as a performance baseline, establishing upper bounds for the evaluation of other algorithms. WOA is a meta-heuristic optimization algorithm designed to effectively balance exploration and exploitation in complex optimization problems.
    \item \textit{Fairness}: This algorithm focuses on fair resource allocation, ensuring that agents receive proportionate spectrum and beamforming resources for transmission in each time slot, promoting equitable access and utilization across the network.
    
\end{itemize}
\begin{figure}[t]
        \centering
        \captionsetup{justification=centering,singlelinecheck=false}
        \includegraphics[width=0.8\columnwidth]{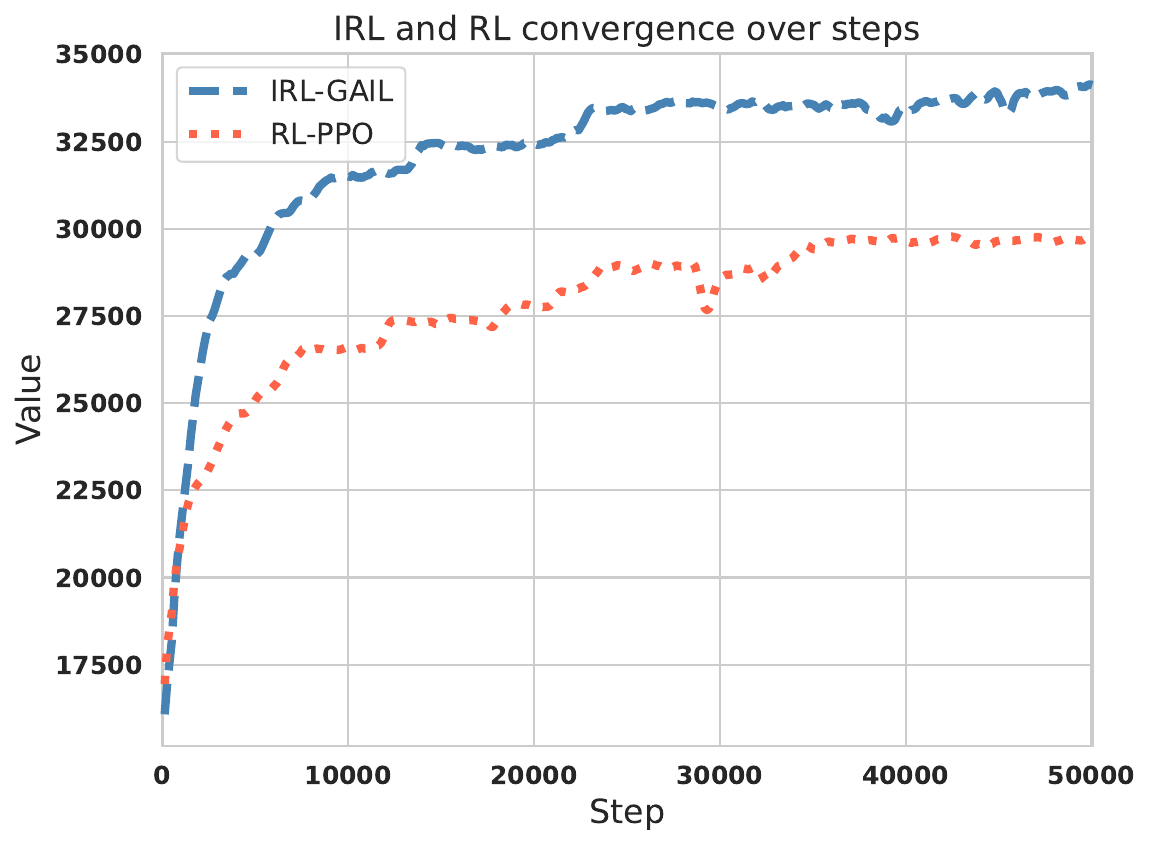}
        \caption{Comparison of the convergence of the proposed IRL algorithm with the SOTA RL-PPO algorithm.}
        \label{Convergence}
\end{figure}
\subsection{Analysis of Simulation Performance}
In Fig. \ref{Convergence}, we compare the training convergence of the proposed \textit{IRL-GAIL} approach with the established \textit{RL-PPO} algorithm. The results indicate that \textit{IRL-GAIL} outperforms \textit{RL-PPO} in both convergence speed and final reward values. After convergence, \textit{IRL-GAIL} achieves reward values $14.6$ units higher than \textit{RL-PPO}. The better performance can be attributed to \textit{IRL-GAIL}'s improved ability to handle large state and action spaces. The faster convergence of \textit{IRL-GAIL} is particularly significant for applications in dynamic networks where conditions change rapidly. The performance gap observed may be partially due to the limited network size used in this experiment, which potentially constrained \textit{RL-PPO}'s optimization capabilities under the given conditions. These findings suggest that \textit{IRL-GAIL} offers advantages over traditional RL approaches, particularly in complex environments with large state and action spaces. However, further research is needed to fully understand the factors contributing to this performance difference.

\begin{figure}[t]
        \centering
        \captionsetup{justification=centering,singlelinecheck=false}
        \includegraphics[width=0.8\columnwidth]{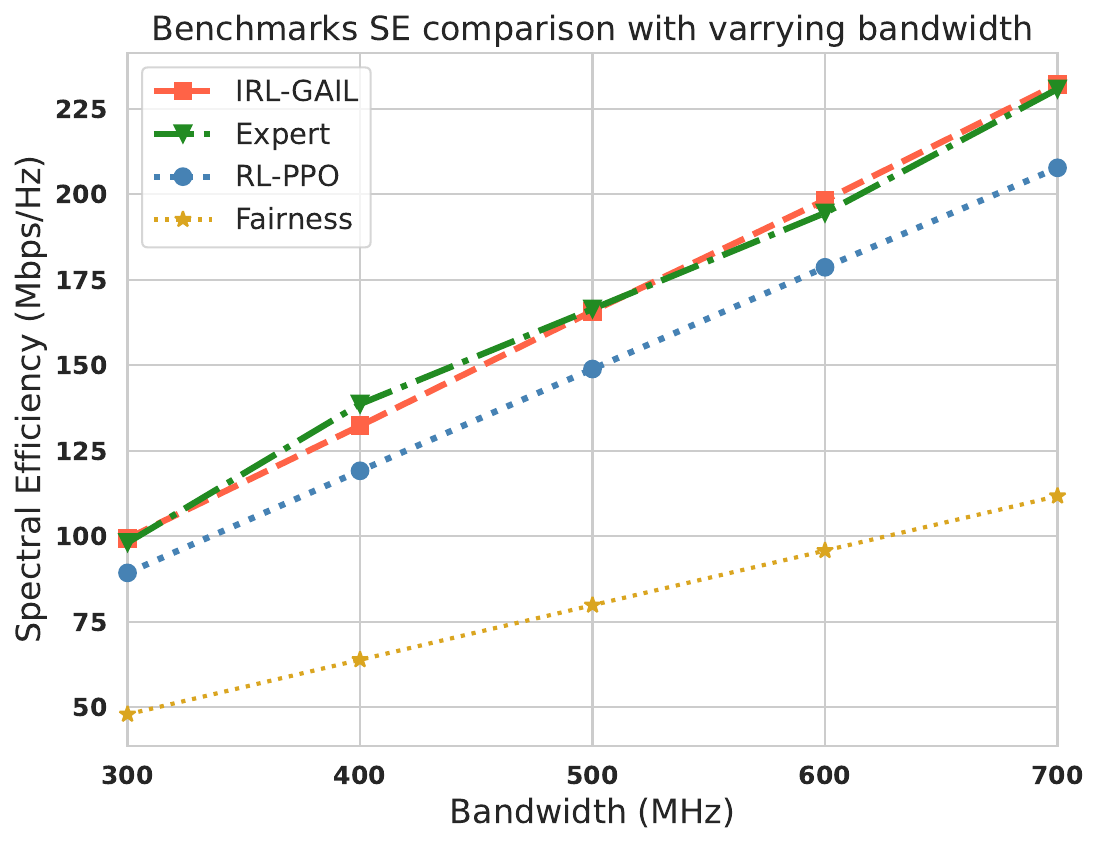}
        \caption{Comparison of the total varying bandwidths of the proposed IRL algorithm with benchmarks.}
        \label{varying_BW_SE}
\end{figure}
Fig. \ref{varying_BW_SE} illustrates the SE performance of the proposed \textit{IRL-GAIL} algorithm compared to other benchmark algorithms, including \textit{RL-PPO}, \textit{Expert}, and \textit{Fairness}. The results show a clear trend: as the satellite bandwidth increases, the total SE improves for all algorithms. Notably, the proposed \textit{IRL-GAIL} algorithm performs nearly as well as the \textit{Expert}, which serves as the upper bound for performance evaluation. Furthermore, \textit{IRL-GAIL} outperforms \textit{RL-PPO} by $10.52\%$ and achieves a $51.87\%$ higher SE than the \textit{Fairness} algorithm. These findings confirm that the proposed \textit{IRL-GAIL} algorithm is highly effective in enhancing SE performance, even under varying spectrum bandwidth conditions. The better performance reinforces the adaptability and robustness of \textit{IRL-GAIL} in dynamic environments. 

\begin{figure}[t]
        \centering
        \captionsetup{justification=centering,singlelinecheck=false}
        \includegraphics[width=0.8\columnwidth]{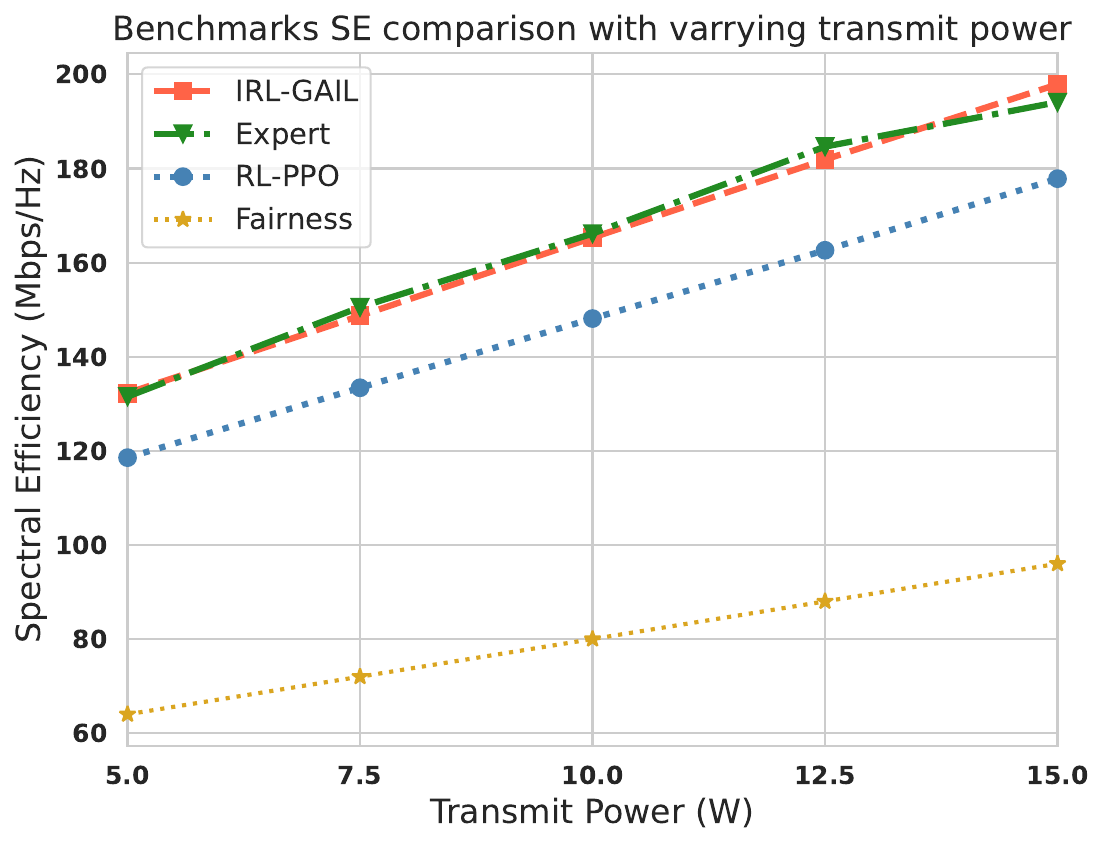}
        \caption{Comparison of the total varying beamforming power of the proposed IRL algorithm with benchmarks.}
        \label{varying_TP_SE}
\end{figure}
\begin{figure}[t]
\centering
\captionsetup{justification=centering,singlelinecheck=false}
\includegraphics[width=0.8\columnwidth]{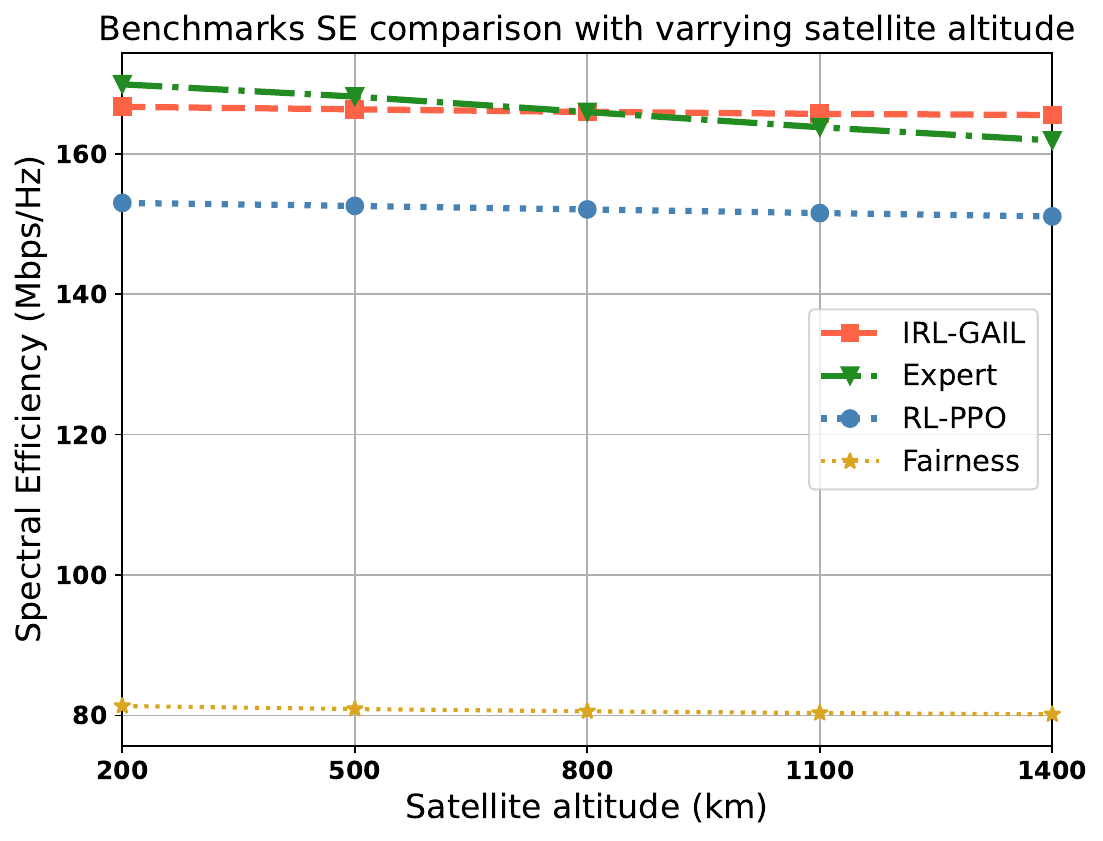}
\caption{Comparison of the total varying satellite altitude of the proposed IRL algorithm with benchmarks.}
\label{varying_altitude_SE}
\end{figure}

Fig. \ref{varying_TP_SE} presents a detailed comparison of the SE performance of the proposed \textit{IRL-GAIL} algorithm against several benchmark algorithms, including \textit{RL-PPO}, \textit{Expert}, and \textit{Fairness}. The results demonstrate a consistent pattern, i.e., all algorithms exhibit improved SE as the satellite's beamforming transmit power increases. However, the proposed \textit{IRL-GAIL} algorithm demonstrates a remarkable ability to achieve performance close to the \textit{Expert} benchmark, which serves as the theoretical upper bound for this evaluation. More significantly, \textit{IRL-GAIL} outperforms \textit{RL-PPO} by $10.13\%$ and the \textit{Fairness} algorithm by an impressive $51.48\%$ in SE performance. These findings demonstrate \textit{IRL-GAIL}'s better efficiency in optimizing SE under varying beamforming transmit power conditions. Additionally, they reinforce the algorithm's scalability and robustness in dynamic network environments. This validates that \textit{IRL-GAIL} is particularly well-suited for real-world applications where adaptive performance for SE optimizing is crucial. The algorithm demonstrates a balance between efficiency and adaptability, making it well-suited for modern satellite communication systems where optimal spectral efficiency is crucial for maximizing resource use.

Fig. \ref{varying_altitude_SE} provides a comparative analysis of the SE performance of the proposed \textit{IRL-GAIL} algorithm alongside established benchmarks, such as \textit{RL-PPO}, \textit{Expert}, and \textit{Fairness}. The results demonstrate a consistent pattern, i.e., SE decreases slightly for all algorithms as satellite altitude increases. This trend is primarily attributable to the greater path losses that occur over longer transmission distances. Despite the challenge, the \textit{IRL-GAIL} algorithm impressively showcases its capability to closely match the performance of the \textit{Expert} benchmark, representing the theoretical upper limit in this context. Moreover, \textit{IRL-GAIL} achieves an SE improvement of $8.72\%$ over \textit{RL-PPO} and an impressive $51.58\%$ improvement over the \textit{Fairness} algorithm. These results reinforce the better efficiency of \textit{IRL-GAIL} in optimizing SE across varying satellite altitudes. Additionally, the algorithm demonstrates robust scalability and adaptability, crucial for dynamic network environments. This suggests that \textit{IRL-GAIL} is highly suitable for real-world applications, particularly in scenarios where adaptive performance under changing conditions is essential. The algorithm’s ability to maintain high efficiency, even under the challenges of increased altitude, emphasizes its potential to meet the demands of modern satellite communication systems, where maximizing spectral efficiency is critical for optimal resource utilization.

\begin{figure}[t]
        \centering
        \captionsetup{justification=centering,singlelinecheck=false}
        \includegraphics[width=0.9\columnwidth]{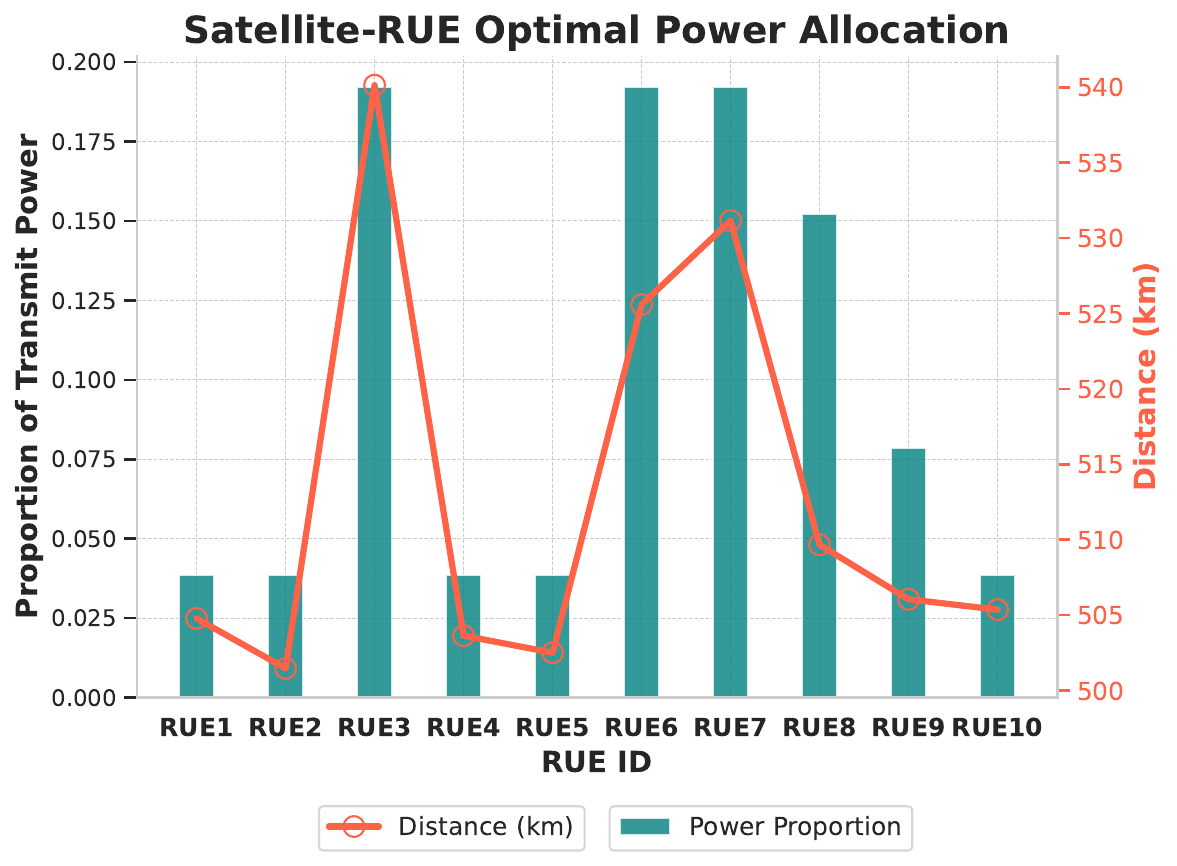}
        \caption{Satellite optimal proportion of transmit beamforming power allocation for RUEs' cluster using the \textit{IRL-GAIL}}
        \label{power_allocation}
\end{figure}
Fig. \ref{power_allocation} illustrates how the proposed \textit{IRL-GAIL} algorithm enables a single satellite to effectively manage beamforming transmit power allocation optimization variable for each associated RUE for each time slot. From these results, we can see that the \textit{IRL-GAIL} algorithm dynamically adjusts power allocation based on the varying distances of RUEs from the satellite. In particular, as the distance between the satellite and a given RUE increases, the algorithm proportionally allocates more power to that RUE to compensate for the higher path loss. Conversely, for RUEs closer to the satellite, the algorithm allocates less power to maintain the optimal SE. This dynamic power control highlights the algorithm’s ability to optimize resource allocation in response to changing network conditions, ensuring efficient power usage and maintaining communication quality for all RUEs.

\begin{figure}[t]
        \centering
        \captionsetup{justification=centering,singlelinecheck=false}
        \includegraphics[width=0.9\columnwidth]{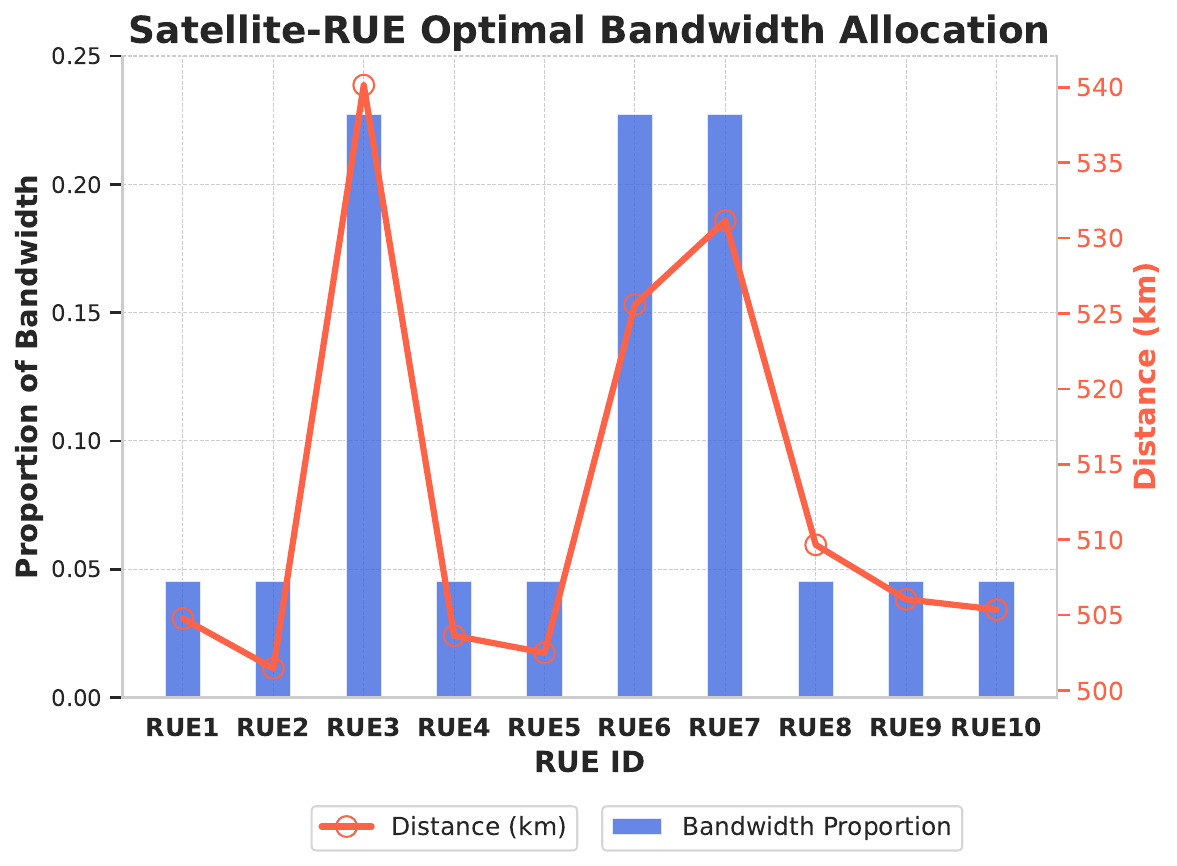}
        \caption{Satellite optimal proportion of bandwidth allocation for RUEs' cluster using the \textit{IRL-GAIL}}
        \label{bandwidth_allocation}
\end{figure}
\begin{figure*}[t]
    \centering
    \begin{subfigure}[t]{0.48\textwidth}
        \centering
        \captionsetup{justification=centering,singlelinecheck=false}
        \includegraphics[width=0.9\columnwidth]{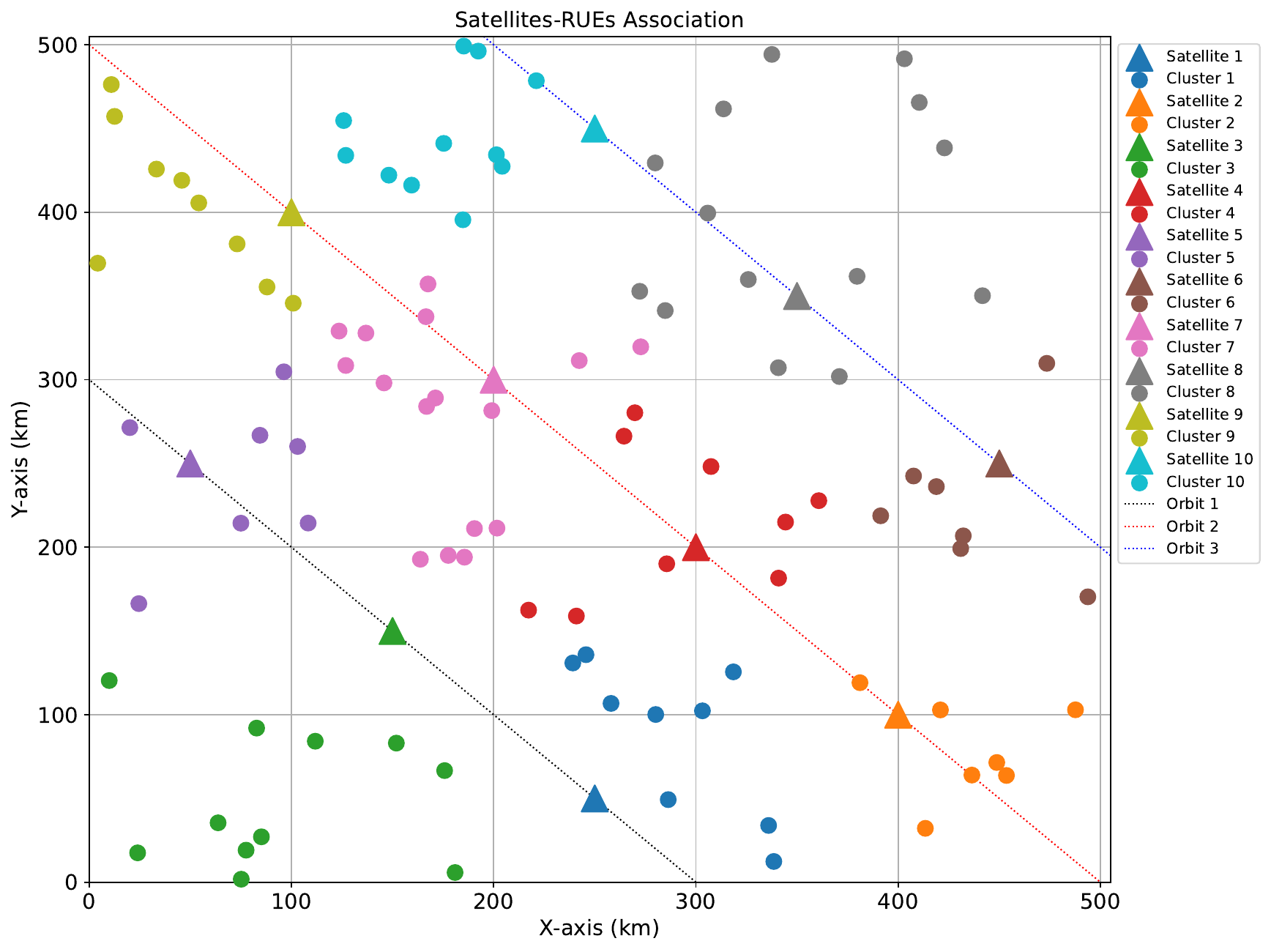}
        \caption{Satellite-RUEs cluster association snippet at time slot = $1$.}
        \label{Satellite-RUE-cluster-association_1}
    \end{subfigure}
    \hfill
    \begin{subfigure}[t]{0.48\textwidth}
        \centering
        \captionsetup{justification=centering,singlelinecheck=false}
        \includegraphics[width=0.9\columnwidth]{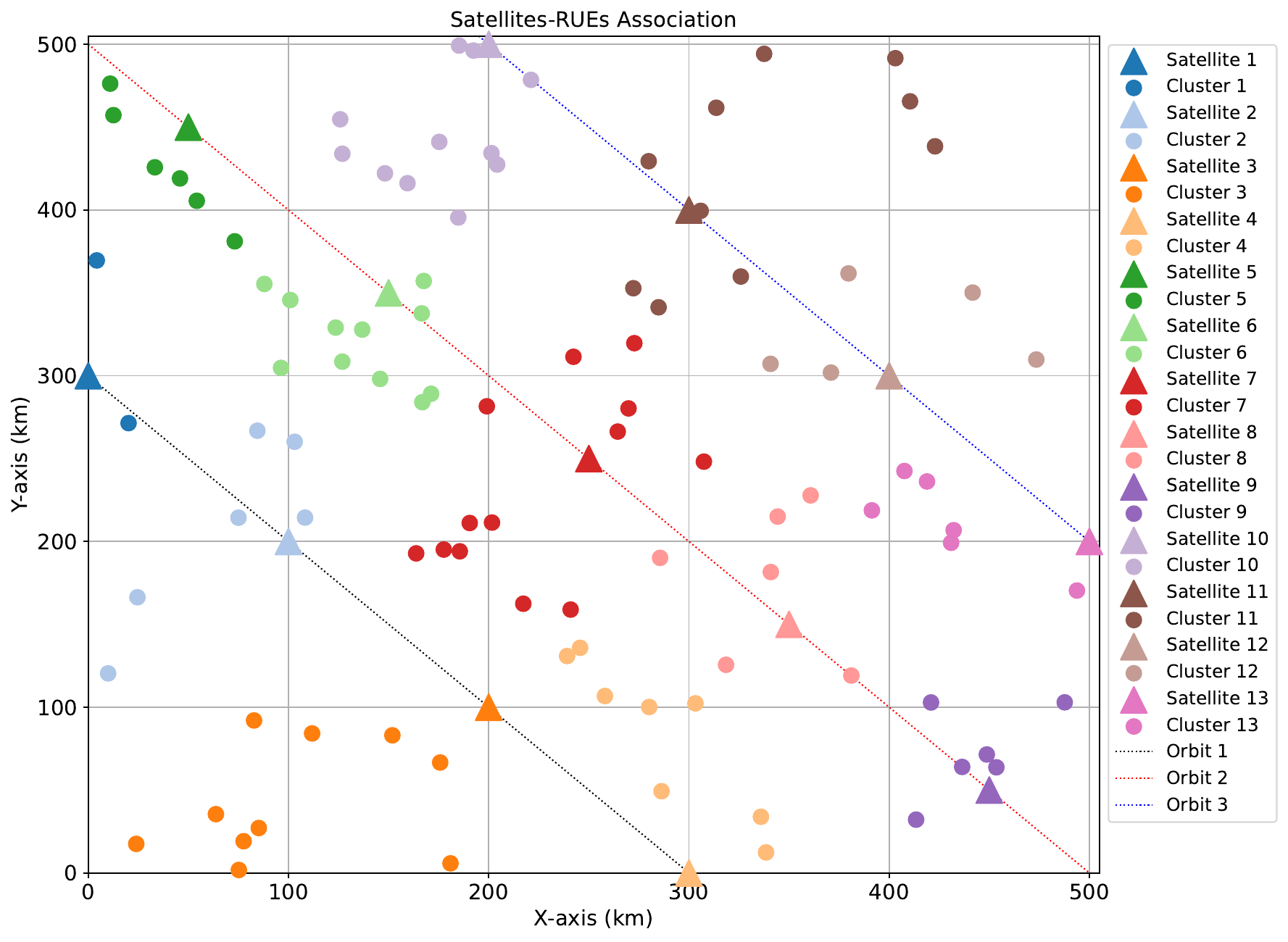}
        \caption{Satellite-RUEs cluster association snippet at time slot = $2$.}
        \label{Satellite-RUE-cluster-association_2}
    \end{subfigure}
    \caption{Walker constellation-based satellite associations with their respective RUE clusters in 2D.}
     \label{Satellite-RUE-cluster-association}
\end{figure*}

Similarly, Fig. \ref{bandwidth_allocation} demonstrates the effectiveness of the proposed \textit{IRL-GAIL} algorithm in managing dynamic bandwidth spectrum allocation optimization variable for each RUE within a satellite communication system. The results showcase that \textit{IRL-GAIL} adjusts spectrum allocation in real-time based on the varying distances between RUEs and the satellite. Specifically, the algorithm allocates more spectrum to RUEs that are farther from the satellite to compensate for increased path loss, thereby maintaining optimal SE. Conversely, for RUEs located closer to the satellite, the algorithm allocates less spectrum, as less compensation is needed to sustain optimal SE. This dynamic adjustment of spectrum allocation reinforces the algorithm’s ability to adapt to changing network conditions, optimizing resource utilization while ensuring consistent communication quality for all RUEs. The ability of \textit{IRL-GAIL} to efficiently balance bandwidth allocation based on the specific needs of each RUE highlights its potential for improving the overall performance and scalability of satellite communication systems in real-world scenarios.

Fig. \ref{Satellite-RUE-cluster-association} illustrates the association between satellites and their respective optimal clusters of Remote User Equipment (RUEs). Specifically, Figures \ref{Satellite-RUE-cluster-association_1} and \ref{Satellite-RUE-cluster-association_2} showcase two snapshots of these optimal satellite-RUE cluster associations over time. The satellites follow a Walker constellation, which maps their movement onto a 2D representation of the Earth's surface. The area of interest for this experiment covers a $500 \times 500~$km$^2$ region. The results demonstrate that the proposed algorithm, based on a matching game approach, dynamically manages the association of RUE clusters to satellites in orbit. This dynamic association is driven by an optimal utility function defined in Equation \ref{eq_pre_ap}, which ensures that the satellite constellation continuously adjusts to changing conditions. Importantly, when a new satellite enters the desired area, the algorithm creates new connections between satellites and RUEs and reorganizes the RUE clusters to ensure the best possible SE. These results show that the algorithm is effective and can adapt to different situations. By constantly adjusting the connections between satellites and RUEs based on satellite movement and network changes, the algorithm makes sure that SE remains at its highest level throughout the association window. This shows that the algorithm can improve the performance of satellite communication networks in real-world situations, where being able to adapt is essential for efficient and reliable communication.

\section{Conclusion}
\label{conc}
In this research, we explored the problem of optimizing SE in multi-satellite systems. We focused on the challenges of beamforming, spectrum allocation, and RUE association. Traditional RL methods often rely on manually designed reward functions, which can be biased and require extensive tuning. To address these issues, we introduced a novel IRL approach based on the GAIL framework. Our method automatically learns reward functions, eliminating the need for manual design. We formulated an optimization problem to maximize SE while ensuring RUEs' QoS. This complex problem was solved using a combination of the many-to-one matching theory and the MA-AFIRL framework. This decentralized approach allowed agents to learn optimal policies efficiently through interactions with the environment, improving scalability and performance. The WOA acted as an expert policy, providing data to train the automatic reward function in the GAIL framework. Our simulation results demonstrate the effectiveness of our proposed MA-AFIRL method, which significantly outperforms traditional RL methods like PPO, achieving up to a $14.6\%$ improvement in both convergence speed and reward performance. These findings establish a new benchmark for NTN optimization, highlighting the potential of GAIL-powered AFIRL for more autonomous and intelligent network management solutions. This work paves the way for future advancements in NTN optimization and wireless communication.





\bibliographystyle{IEEEtran}
\bibliography{ref}

\end{document}